\documentclass[11pt, letterpaper]{article}
\usepackage{epsfig,mst-stylefile,amssymb,amsmath, multirow, url, tcolorbox,booktabs, enumitem}

\usepackage{natbib}
\usepackage[utf8]{inputenc}
\usepackage{bbm}
\usepackage{hyperref}

\usepackage[export]{adjustbox}
\usepackage{tikz,pgfplots}

\usepackage{amsmath}

\newtheorem{@assumption}{\sc Assumption}[section]

\newcommand{\calN}{{\mathcal N}}

\newcommand{\bbE}{{\mathbb{E}}}

\newcommand{\bbR}{{\Bbb R}}
\newcommand{\bbN}{{\Bbb N}}

\newcommand{\bbP}{{\mathbb{P}}}

\newcommand{\E}{\mathbb{E}}

\newcommand {\var}{\qopname\relax n{\textrm{Var}}}

\newcommand{\fxhat}{\widehat f_X(x)}
\newcommand{\fyhat}{\widehat{f}_Y(y)}
\newcommand{\fyhatm}{\widehat{f}^{(m)}_Y(y)}
\newcommand{\fy}{f_Y(y)}
\newcommand{\fx}{f_X(x)}
\newcommand{\gx}{g_X(x)}
\newcommand{\px}{p_X(x)}
\newcommand{\gpd}{g_{\xi,\beta}(u)}
\newcommand{\PL}{\bbP(X\le x_L)}
\newcommand{\PR}{\bbP(X \ge x_R)}
\newcommand{\Prr}{\bbP(X > x_R)}
\newcommand{\sL}{x\le x_L}
\newcommand{\sR}{x\ge x_R}
\newcommand{\sM}{x_L < x< x_R}
\newcommand{\Xio}{X_{0,i}}
\newcommand{\yR}{ y \ge y_R}
\newcommand{\yM}{y_L < y < y_R}
\newcommand{\yL}{y \le y_L}
\newcommand{\ylyr}{(y_L, y_R)}
\newcommand{\xlxr}{(x_L,x_R)}
\newcommand{\bbone}{{\mathbbm{1}}}
\newcommand{\GP}{\mathcal{G}\mathcal{P}}
\newcommand {\hxi}{ \widehat{\xi}}
\newcommand {\hdel}{ \widehat{\beta}}
\newcommand {\gfunction}{\displaystyle \frac{1}{\beta}(1+\frac{\xi u}{\beta})^{-\frac{1}{\xi}-1}}

\newcommand {\FbhR}{\widehat{\bar F}_X(x_R)}
\newcommand {\Xin}{x_{(j)}}
\newcommand {\Xiin}{x_{(j+1)}}
\newcommand {\Yiin}{y_{(j+1)}}
\newcommand {\Yin}{y_{(j)}}
\newcommand{\xRest}{X_{0,(N_0-r+1):N_0}}
\newcommand{\xiNr}{\xi_{N_0,r}}
\usepackage{subcaption}
\usepackage{algorithm}
\usepackage{algpseudocode}
\algrenewcommand\alglinenumber[1]{\tiny #1:}

\begin{document}
\title{Sampling low-fidelity outputs for estimation \\
of high-fidelity density and its tails  }

\author{
\begin{tabular}{cc}
Minji Kim & Kevin O'Connor \\
University of North Carolina & Optiver \\
\end{tabular}
\and 
\begin{tabular}{cc}
Vladas Pipiras & Themistoklis Sapsis \\
University of North Carolina & MIT
\end{tabular}
}


\date{\today}
\maketitle

\begin{abstract}
In a multi-fidelity setting, data are available under the same conditions from two (or more) sources, e.g.\ computer codes, one being lower-fidelity but computationally cheaper, and the other higher-fidelity and more expensive. This work studies for which low-fidelity outputs, one should obtain high-fidelity outputs, if the goal is to estimate the probability density function of the latter, especially when it comes to the distribution tails and extremes. It is suggested to approach this problem from the perspective of the importance sampling of low-fidelity outputs according to some proposal distribution, combined with special considerations for the distribution tails based on extreme value theory.
The notion of an optimal proposal distribution is introduced and investigated, in both theory and simulations. The approach is motivated and illustrated with an application to estimate the probability density function of record extremes of ship motions, obtained through two computer codes of different fidelities.
\end{abstract}

%


\section{Introduction}
\label{s:introduction}
In modeling physical phenomena, it is common to have several models of varying fidelity and computational cost, with higher fidelity associated with greater cost. In such multi-fidelity (MF, for short) settings, there has been considerable effort by many researchers recently on how questions about a high-fidelity output $Y$ can exploit information about the corresponding low-fidelity output $X$. \cite{peherstorfer2018survey} categorizes MF strategies into three types, each applicable across various objectives: ``adaptation", where high-fidelity information is used to enhance the lower-fidelity model (\cite{Kennedy:Hagan:2000}, \cite{kim2023calibration}); ``fusion", which involves the combined use of multiple models (\cite{gorodetsky2020}, \cite{peherstorfer:2019:mfmc}, \cite{pham2022ensemble}); and ``filtering", where the low-fidelity model is explored to determine where to evaluate the high-fidelity model (\cite{Narayan2014stochastic}, \cite{pham2022ensemble}, \cite{peherstorfer2016multifidelity}).
Some studies, including some cited, can fall under multiple types. This paper considers the following problem of the filtering and fusion types, described next starting with our motivation.

The application of our interest concerns modeling ship motions (and especially their extremes) in Naval Architecture. The motions are considered for a ship in irregular (random) waves, and will be driven by a random wave height $\zeta(t,x)$ at time $t$ and location $x$. Assuming for simplicity the case of head or following waves with one-dimensional $x$, the commonly used Longuet-Higgins model postulates that
\begin{equation}
    \label{e:intro:Longuet-height}
    \zeta(t,x) = \sum_{n=1}^{N_w} a_n \cos{(k_n x - w_n t + \varepsilon_n)}, 
\end{equation}
where $w_n>0$ form a set of typically equally spaced frequencies, $k_n$ are the so-called wave numbers (e.g., $k_n=w_n^2/g$ in deep water with the gravitational acceleration constant $g$), and $a_n>0$ are deterministic amplitudes (expressed through some spectrum function evaluated at $w_n$). The number of frequencies $N_w$ is in the order of a few hundreds. The only {\it random} components in \eqref{e:intro:Longuet-height} are the so-called random phases $\varepsilon_n$ taken as independent and uniformly distributed random variables on $(0,2\pi)$. See \cite{LH1957} and \cite{lewis1989principles}. In computer experiments, simulations are run with \eqref{e:intro:Longuet-height} for records of certain length in time $t$ (e.g., 30 minutes). Each record is thus associated with a particular set of random phases $\{\varepsilon_n\}_{n=1}^{N_w}$, or a particular random seed $\omega$ (not to be confused with frequencies $w_n$) used to generate $\varepsilon_n = \varepsilon_n(\omega), ~n=1,\dots,N_w$.
Random seeds can also be thought (and relabeled) as record numbers, i.e., 1, 2, 3 and so on.

Given the same random excitation \eqref{e:intro:Longuet-height} associated with some random seed or record number, researchers in Naval Architecture use a range of computer codes to generate ship motions and related quantities. For example, two such codes to be referred to below are SimpleCode (SC; \cite{weems:wundrow:2013}) and Large Amplitude Motion Program (LAMP; \cite{lin:yue:1991}, \cite{shin2003nonlinear}). The two codes differ in the underlying physics which is not relevant for this work, with LAMP being higher fidelity. At the same time, SC is computationally more efficient: if a 30-minute record of ship motions takes about 2-3 seconds to generate for SC, this time could be 15-20 minutes or longer for LAMP depending on what outputs are sought.

\begin{figure}
\centerline{
\includegraphics[width=2.5in,height=2.1in]{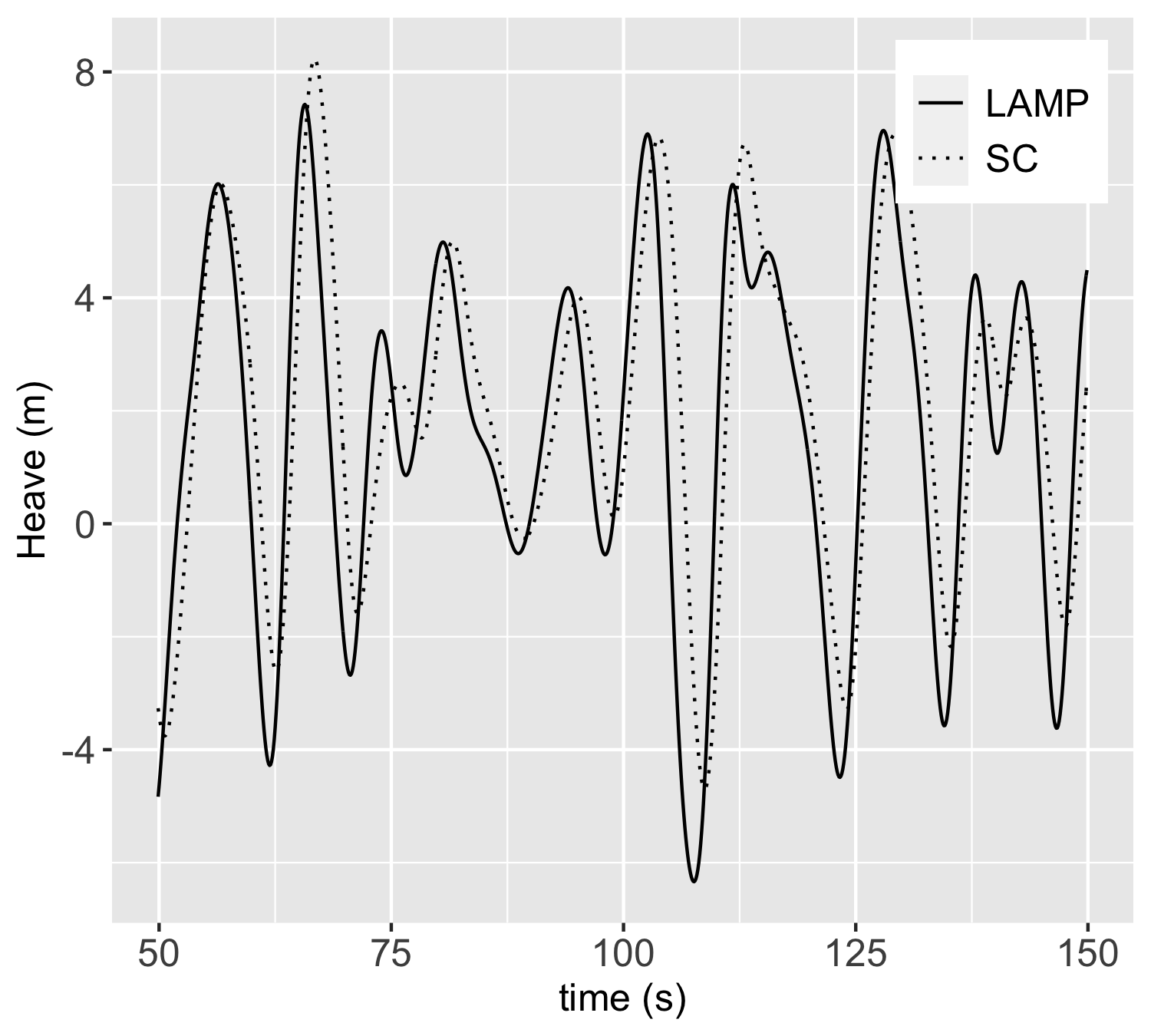}
\includegraphics[width=2.5in,height=2.1in]{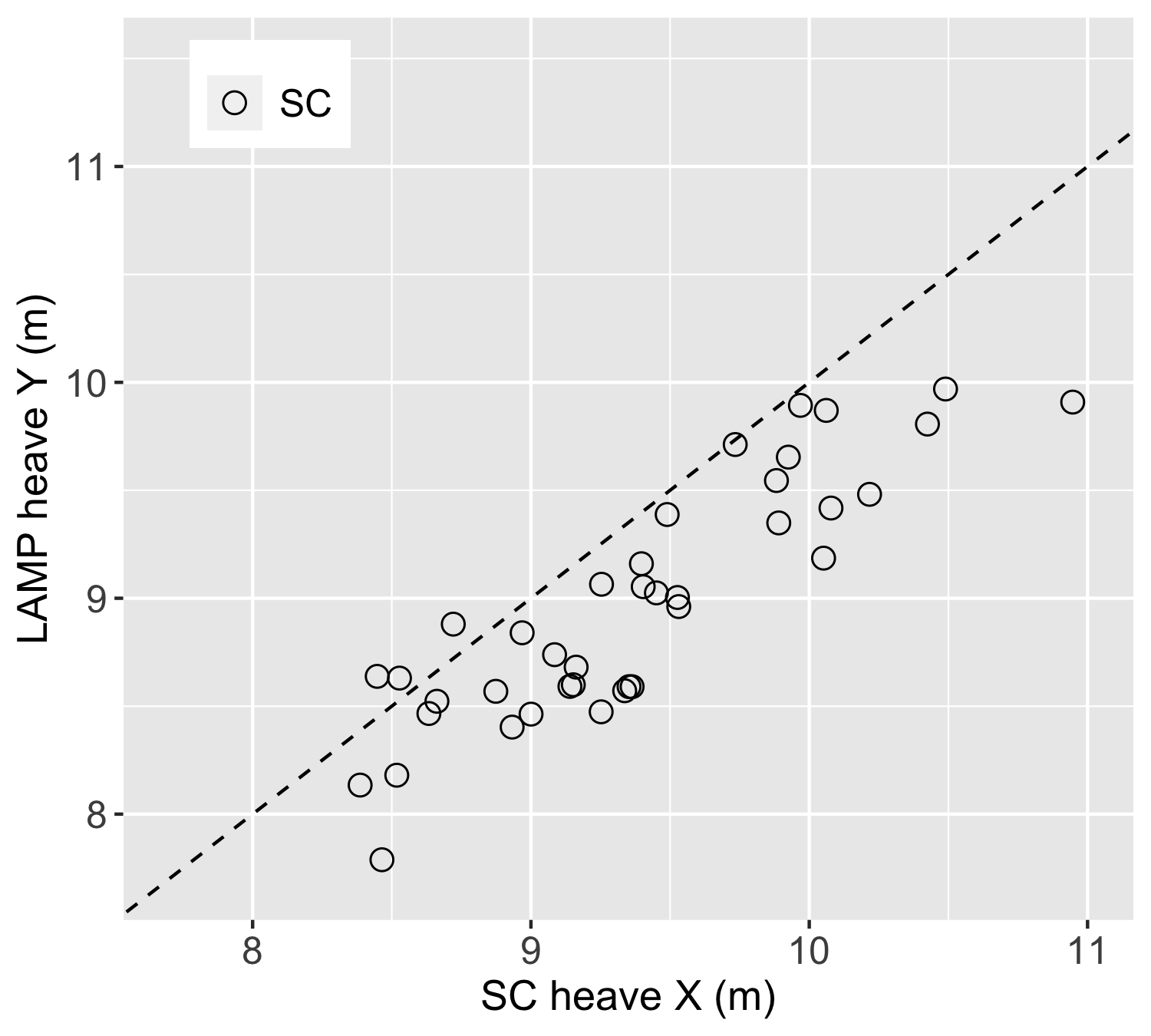}
}
\caption{Left: Heave motion for LAMP and SC. Right: LAMP versus SC heave record maxima. The dashed line is the $45^\circ$ line.}
\label{f:intro-heave}
\end{figure}
The point to keep in mind is that LAMP and SC will generate different ship motions even for the same excitation \eqref{e:intro:Longuet-height} or record number (random seed). Figure~\ref{f:intro-heave}, left plot, illustrates this for one of the ship motions, heave, over a time window of 100 secs; this is for a particular ship hull geometry, wave conditions, heading, and so on. Figure~\ref{f:intro-heave}, right plot, presents the scatterplot of heave maxima of LAMP and SC for 20 randomly selected records of 30-minute length each. Such data for many records could be used to make statements about the occurrence of extremes.
Viewing this as a MF setting discussed above, the high-fidelity output $Y = Y(\omega)$ is the LAMP record heave maximum, and the low-fidelity output $X = X(\omega)$ is the SC record heave maximum, where $\omega$ is the same record number (random seed). The data presented in Figure~\ref{f:intro-heave}, right plot, would be denoted $(X_1, Y_1), \dots, (X_{20}, Y_{20})$, where $(X_i, Y_i)$ are i.i.d.\ copies of $(X,Y)$. 

\begin{figure}
\centerline{
\includegraphics[width=2.5in,height=2.1in]
{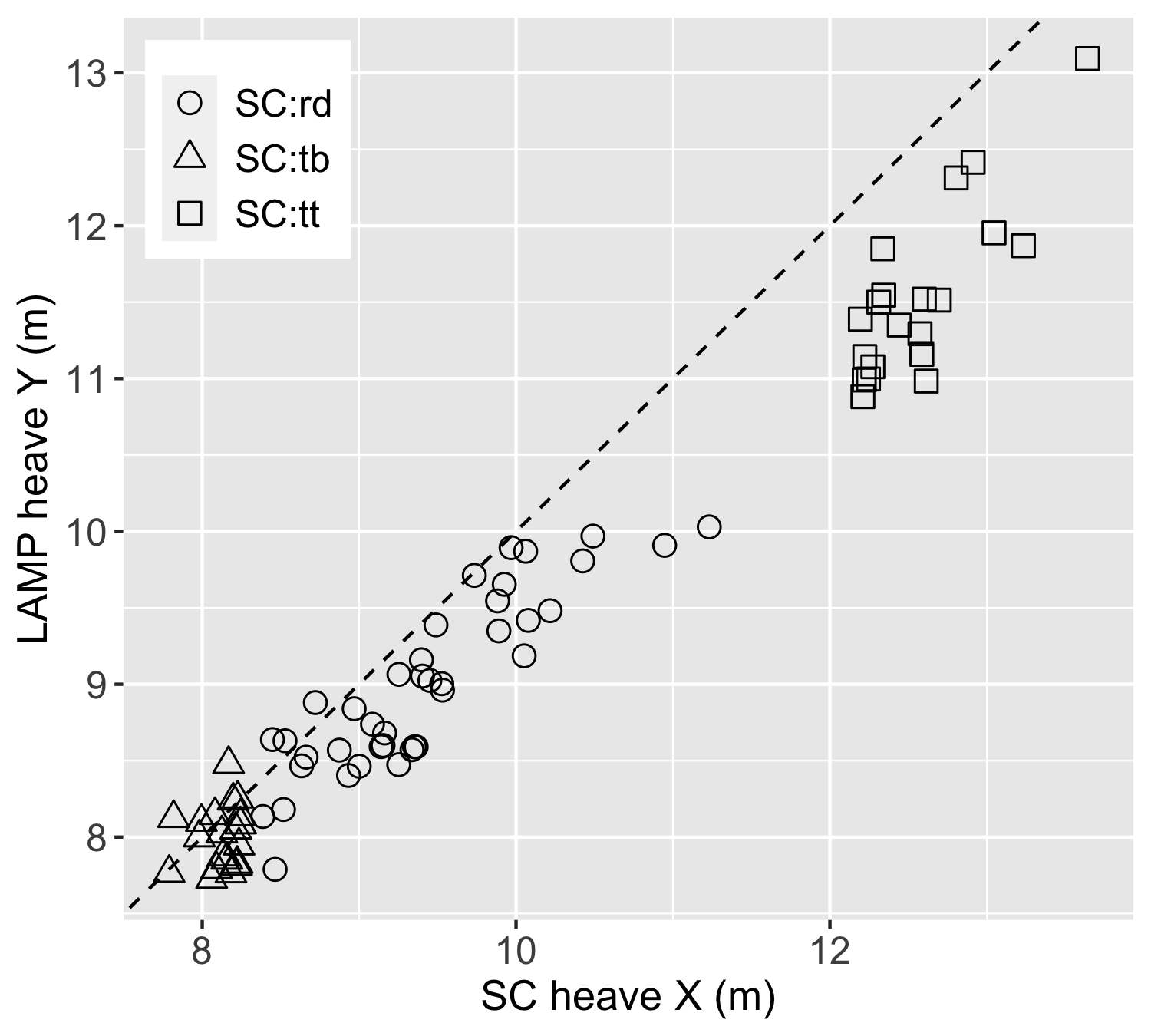}
\includegraphics[width=2.5in,height=2.1in]{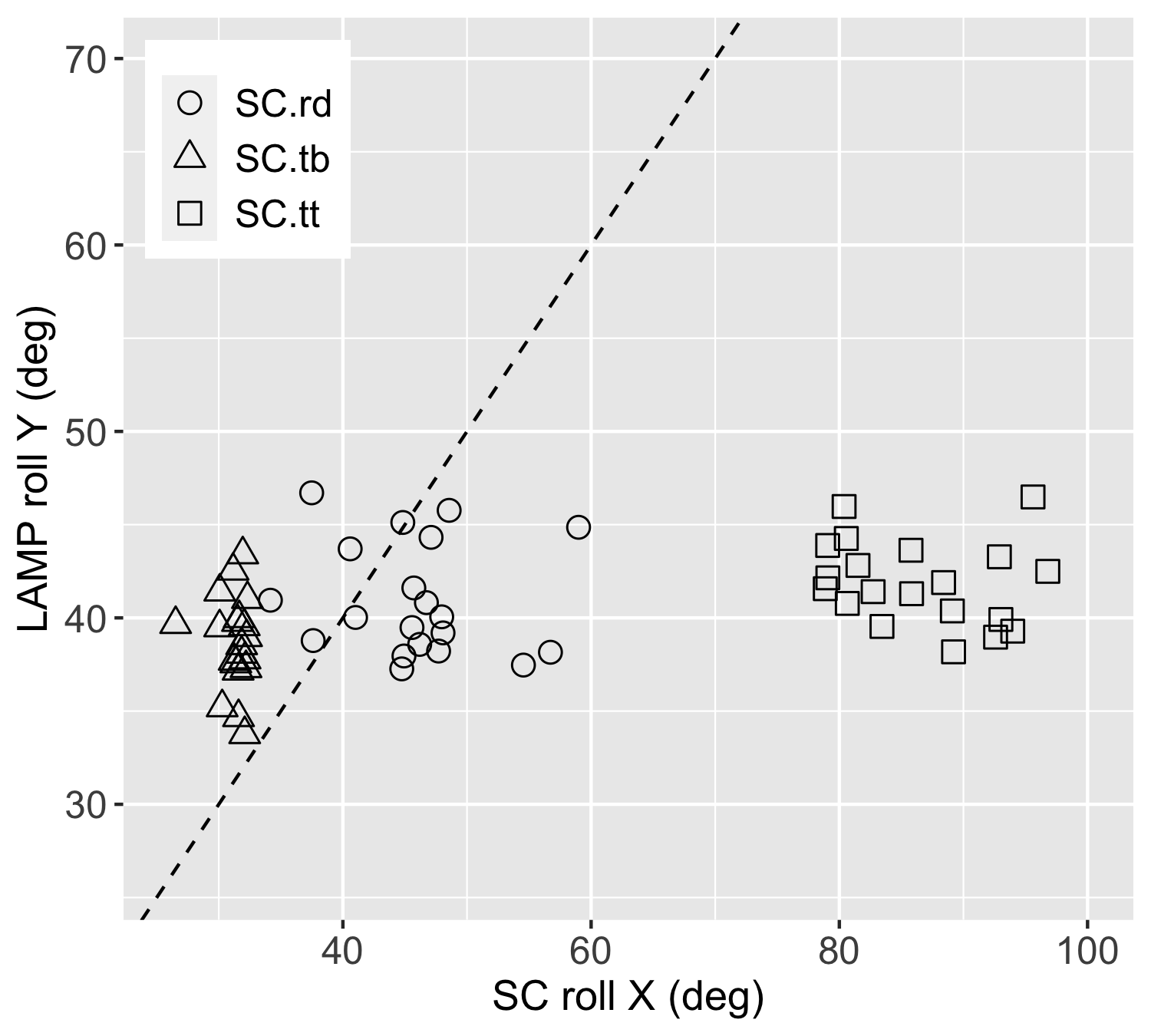}}
\caption{Left: LAMP versus SC heave record maxima including pairs of points corresponding to SC records with 20 top and 20 bottom maxima heave records amongst 2,000. Right: LAMP versus SC roll record maxima including pairs of points corresponding to SC records with 20 top and 20 bottom maxima roll records amongst 2,000.}	
\label{f:intro-sampling}
\end{figure}
In connection to extremes, we are interested in this work to estimate the high-fidelity PDF (probability density function) of $Y$, denoted $\fy$, and especially its tails, leveraging additional data of the low-fidelity output $X$. Furthermore, as it is computationally inexpensive to generate $X$, we assume that we can have much more data for $X$ than for $Y$. 
Importantly, note also that in the application setting described above, $X$ can be generated without the corresponding $Y$. This suggests that we may be more selective for which values of $X = X(\omega)$, and hence their corresponding record numbers (random seeds) $\omega$, we should generate the corresponding expensive values $Y(\omega)$. If the goal is estimating $\fy$ further in the tails (larger and smaller $y$), and if $X$ and $Y$ are strongly correlated, it would make sense to generate $Y$ for larger and smaller observed $X$'s (so that $Y$ would potentially be larger and smaller). This idea is illustrated in Figure~\ref{f:intro-sampling}. The left plot of the figure is a scatterplot akin to the right plot of Figure~\ref{f:intro-heave} and contains the points of the latter under ``rd'' or ``random'' but additional points are added as follows. 2,000 SC records are generated first. Among these, 20 record numbers (and the corresponding random seeds) are identified having 20 largest record heave maxima among the 2,000 SC records. Then, LAMP records are generated for these 20 record numbers and the corresponding LAMP/SC record maxima pairs appear in the scatter plot as the ``tt'' or ``(top) top 20'' points. The ``tb'' or ``(top) bottom 20'' points are obtained similarly but for 20 record numbers with the smallest record maxima among the 2,000 SC records.
Again, since the range of $Y$-values is larger compared to random sampling, such selective sampling (we use a more technical term and method ``importance sampling'' below) should be more advantageous when using lower-fidelity $X$ to inform inference about higher-fidelity $Y$. We also note that such benefit is expected if $X$ and $Y$ are strongly correlated. Strong dependence is not necessarily guaranteed, even for the same record numbers (random seeds). Figure~\ref{f:intro-sampling}, right plot, depicts an analogous scatterplot but for another motion, roll, where the dependence between $X$ and $Y$ is weak.

With our goal and selective sampling procedures discussed above, we are interested in the following questions:
\begin{itemize}
	\item [Q1:] What is an optimal way to sample $X$-outputs (records) and generate the corresponding $Y$-outputs, so that the estimation of the PDF $\fy$ is best? 
	\item [Q2:] For potential sampling schemes, what are the estimators of the PDF $\fy$ in the first place? How does one quantify their statistical uncertainty? 
     \item [Q3:] Should estimation of the PDF be treated separately in the tails, where less (or no) data are available, and how?
\end{itemize}

Though our application is in Naval Architecture, the framework and questions presented above should be of interest in other MF settings where randomness underlies quantities of interest (QoI's). A typical example is a PDE modeling a physical phenomenon with uncertain parameters assumed to be random. A high-fidelity model is a high-fidelity discretization of the PDE, while a low-fidelity model is its approximation, for example, using proper orthogonal decomposition or other surrogate approach. Examples are many in related literature on MF methods for failure probabilities, conditional value at risk (CVaR) and other QoI's. CVaR is considered, for example, in \cite{heinkenschloss2018conditional, heinkenschloss2020adaptive}, where PDEs with random parameters were for a convection-diffusion-reaction problem and a heat problem (for a thermal fin) with certain temperature quantities and their CVaR being of interest. MF estimation of a failure probability is considered, for example, in \cite{peherstorfer2017combining} for the displacement of a cantilever beam assuming its length is modeled by a random variable. As in this work, there is emphasis on distribution tails in failure probability and CVaR. However, we are interested in the whole PDF (rather than some fixed value associated with it) and our sampling methods involve the lower-fidelity values themselves (rather than the underlying random variables, usually low-dimensional). (Some recent work on MF estimation of the whole distribution of QoI's though is available; see \cite{han2023approximate}. Some recent work also considers sampling issues in higher-dimensional setting, as sampling $\varepsilon_n, ~ n=1, \dots, N_w$, in (\ref{e:intro:Longuet-height}) with $N_w$ in a few hundreds; see \cite{Pickering2022}.) 

Another notable MF setting with several computer codes (DNS, RANS) is modeling of turbulent flows. In some instances, one is similarly interested in the uncertainty propagation of random input parameters on turbulence QoI's; see, e.g., \cite{voet2021hybrid}, \cite{Rezaeiravesh2023}. Finally, we note that computer models are also considered with inherent randomness (generating random response for fixed conditions and referred to as stochastic computer models) where our methods could potentially be of interest; see, e.g., \cite{Li2021NonparametricIS}.

To answer the questions Q1--Q3 above, we work in a fairly general framework described in more detail in Section~\ref{s:setting} where we also revisit the questions of interest using its notation and explain key aspects of our approach.
The methods behind our approach are considered in Section~\ref{s:methods}. In Section~\ref{s:gpd}, we extend our discussion to the distribution tail based on the extreme value theory and Section~\ref{s:est} further addresses related issues of sampling and estimation.
Data illustrations, in both simulations and the ship application, can be found in Section~\ref{s:data}. Section~\ref{s:conclusions} concludes.


\section{Setting with notation and key elements of approach}
\label{s:setting}

The setting motivated by the application of Section \ref{s:introduction} is as follows. The (real-valued) variables $X$ and $Y$ will refer to the corresponding lower- and higher-fidelity outputs (e.g.\ motion record maxima for SC and LAMP in our application).
We shall sometimes write ``lo-fi" and ``hi-fi" for lower-fidelity and higher-fidelity.
The variable $(X,Y)$ can be defined as a vector and can be viewed as $X=X(\omega),~ Y=Y(\omega)$, where $\omega$ is a sample point (random seed or record number in our application). Define
\begin{equation}\label{eq:def0}
\begin{split}
f_X(x) : &\text{ PDF of } X \text{ (sampled at random),}\\
f_Y(y) : &\text{ target PDF of } Y \text{ when $X$ follows } f_X(x).
\end{split}
\end{equation}
The PDF $f_Y(y)$ is for $Y$ sampled at random as well, but we describe it as in \eqref{eq:def0} for better comparison below and to follow our application, where for such $X$ (sampled by $f_X(x)$), there is a corresponding value of $Y$.
We refer to $\fy$ as the target PDF because our ultimate goal is its estimation.

\begin{figure}
\centering
\begin{tikzpicture}[xscale=0.85, yscale = 0.65]
    \draw[->] (0,0) -- (11,0) node[right] {lo-fi $X$};
    \draw[->] (0,0) -- (0,10) node[above] {hi-fi $Y$};
    
    \draw (6,-0.1) -- (6,0.1) node[below=2mm] {$\Xio$};
    
    \draw[smooth, thick, domain=2:10.5] plot (\x, {1.5*exp(-0.3*(\x-6)^2)+0.1});
     \node at (9.3,0.7) {$f_X(x)$};
    \draw[smooth, domain=1:10] plot ({1.5*exp(-0.1*(\x-6.4)^2)}, \x-0.5);
     \node at (2.3,7.0) { $f_Y(y)$};
     \node at (1.9,9.6) {$\widehat{f}_Y(y)=?$};
    \draw[gray] (0,0) -- (10,10);

    \foreach \i in {1,2.8, 4, 5, 6, 7, 8,8.7,  10} {
        \draw (\i,-0.1) -- (\i,0.1);
    }
    \foreach \i in {4.5, 4.7, ..., 7.1, 7.5} {
        \draw (\i,-0.1) -- (\i,0.1);
    }
    \foreach \i in {5.6, 5.7, ..., 6.5} {
        \draw (\i,-0.1) -- (\i,0.1);
    }

    \draw[lightgray, dashed] (4,1) -- (4,0);
    \draw[lightgray, dashed] (8,1) -- (8,0);
    \draw[gray] (4,0.8) -- (4,1.2);
    \draw[gray] (8,1.2) -- (8,0.8);
    \draw[gray] (4,1) -- (8,1);
    \node at (8.2,1.7) {\color{darkgray}$p_X(x)$};
    \draw[line width=1pt] (4,-0.2) -- (4,0.2) node[below=3mm] {$x_L$};
    \draw[line width=1pt] (8,-0.2) -- (8,0.2) node[below=3mm] {$x_R$};
    
    \foreach \x/\y in {4.5/4.1, 5.1/5.6, 5.6/5.4, 6.1/6.4, 6.7/6.2, 7.3/7.6, 10/9} {
    \fill (\x,\y) circle (0.1);
    \draw[dashed] (\x,\y) -- (\x,0);
    \draw[dashed] (\x,\y) -- (0,\y);
    }
    \node at (8.3,7) {$(X_i,Y_i)$};
    \node at (11,9.2) {$(X_i,Y_i)$};
    \node at (10.9,4.8) {\color{darkgray}$g_X(x)$};
\end{tikzpicture}
\caption{Illustration of the framework and the notation of this work.}
\label{fig:star}
\end{figure}
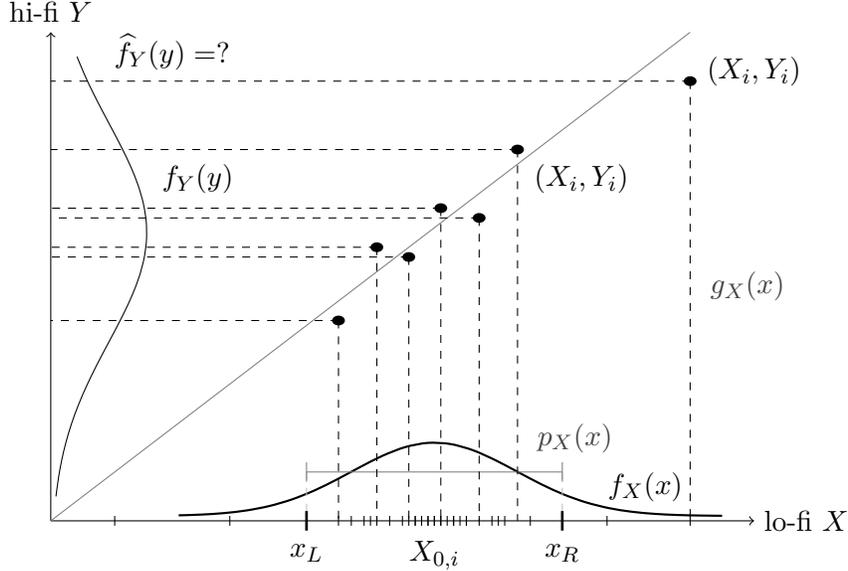
In practice, $f_X(x)$ could be estimated from:
\begin{align}\label{eq:dataX}
    X_{0,1}, \dots, X_{0,N_0} : \text{ data for estimation of } f_X(x).
\end{align}
As $X$ is associated with the less expensive low-fidelity outputs, the data \eqref{eq:dataX} for a large sample size $N_0$ could in principle be generated, without the corresponding outputs of $Y$. In Figure~\ref{f:intro-sampling}, one can think of $N_0=2,000$. In Section~\ref{s:data} with numerical studies, $N_0$ ranges from $10^5$ to around $10^7$. For visual illustration, we will refer to Figure~\ref{fig:star}, where the values of $\Xio$ are marked on the horizontal lo-fi $X$ axis, and a hypothetical PDF $f_X(x)$ from which $\Xio$ are sampled is added to the plot. Naturally, there are more data points (marks) $\Xio$ where the PDF is larger.

The PDF $f_X(x)$ will need to be used in our selective (importance) sampling approach below. Note, however, that from the data \eqref{eq:dataX}, one can expect to estimate $f_X(x)$ well only over: 
\begin{align}\label{eq:xlim}
    (x_L, x_R) : \text{ range of $x$ to estimate } f_X(x).
\end{align}
The larger $N_0$, the larger $(x_L, x_R)$ is expected. We marked this range qualitatively in Figure~\ref{fig:star} as well. We discuss the choice of $x_L, x_R$ in Section~\ref{s:est-t-k}. For simplicity of the argument, we shall suppose henceforth that the PDF $f_X(x)$ can be estimated well enough so that it can be assumed to be known over this range, that is,
\begin{align}\label{eq:xlim2}
    (x_L, x_R) : \text{ range of $x$ to assume } f_X(x) \text{ is {\bf known}.}
\end{align}
The PDF $f_X(x)$ is not assumed to be known outside the range $(x_L, x_R)$.

Over the range $(x_L, x_R)$ in \eqref{eq:xlim}, less data $X_i$ can be resampled according to another, so-called proposal PDF $p_X(x),$ $\sM$. In our application, we think of $X_i$ as sampled from $\{\Xio | \Xio\in (x_L, x_R)\}$. For the selected $X_i$, the corresponding values of $Y_i$ can be obtained. In Figure~\ref{fig:star}, we depict a uniform PDF $p_X(x)$ and a few points $(X_i, Y_i)$ sampled from this selective (importance) scheme. Summarizing, we have
\begin{equation}
    \label{eq:def1}
    \begin{split}
        &p_X(x) : \text{ proposal PDF for } x \in (x_L, x_R),\\
        &X_i\in(x_L,x_R) : \text{ sampled from $\Xio$ according to } p_X(x),\\
        &Y_i : \text{ the corresponding $Y$-values of } X_i.
    \end{split}
\end{equation}
Again, the number of $X_i$'s should be much smaller than $N_0$, since hi-fi $Y_i$ are now generated as well. The purpose of $p_X(x)$ is to resample fewer $\Xio$'s while still covering the observed range $\xlxr$.

We would like to use the data $Y_i$ to estimate the target PDF $f_Y(y)$. The PDF is depicted in Figure~\ref{fig:star} along the hi-fi $Y$-axis, with the question of what estimator $\widehat{f}_Y(y)$ to take indicated as well. In the approach taken below, we will effectively rely on a well-known and widely used kernel-based PDF estimator with suitable importance weights. It will be important that $\fx$ is assumed to be known for $x \in \xlxr$ as in \eqref{eq:xlim2}, since both $\px$ and $\fx$ will define the importance weights for $x\in \xlxr$; see \eqref{e:density-ks}--\eqref{e:density-ks-weights} below.

There is one additional important element that we want to bring to the discussion above. Note that we thus far excluded from our discussion any outputs $\Xio \ge x_R$ or $\Xio \le x_L$. These outputs, however, potentially carry a very useful information about extremes of $X$ and, if $X$ and $Y$ are strongly dependent, also about extremes of $Y$. See Figure~\ref{f:intro-sampling}. In fact, we would like to work with a proposal PDF $g_X(x)$ that samples (ideally all) extreme outputs $\Xio$. Such density will be constructed in Section~\ref{s:methods} below. It is noted in Figure~\ref{fig:star} along one point $(X_i,Y_i)$ with the largest $X_i$. Summarizing and introducing another notation:
\begin{equation}
    \label{eq:def2}
    \begin{split}
        &g_X(x) : \text{ proposal PDF for the whole range of } x,\\
        &X_1,\dots,X_N : \text{ sampled from $\Xio$ according to $g_X(x)$, including extremes of } \Xio,\\
        &g_Y(y) : \text{ PDF of $Y$ when $X$ follows } g_X(x).
    \end{split}
\end{equation}
Again, we think of $N$ in \eqref{eq:def2} as being much smaller than $N_0$ in \eqref{eq:dataX}.

With the introduced notation, the questions of Section~\ref{s:introduction} can be rephrased as:
\begin{itemize}
	\item [Q1:] What $\px$ should be taken? Is there an optimal way to do so?
	\item [Q2:] How is the estimator $\fyhat$ of $\fy$ defined?
    \item [Q3:] What is the difference between $\px$ and $\gx$? How are the tails of $\fy$ estimated?
\end{itemize}
We address these questions in Sections~\ref{s:methods} and \ref{s:gpd} below. In some of our developments in Section~\ref{s:methods}, we shall assume that $X$ and $Y$ are related through one of the following cases:
\begin{eqnarray}
	\mbox{Homoscedastic} & : & Y = m(X) + \sigma \epsilon, \label{e:setting-1} \\
	\mbox{Heteroscedastic} & : & Y = m(X) + \sigma(X) \epsilon, \label{e:setting-2}
\end{eqnarray}
where $\epsilon$ has mean $0$, variance $1$ and is independent of $X$. The most general bivariate relationship between $(X,Y)$ can be expressed as $Y = m(X) + \eta$ with $m(X) = \E(Y|X)$ and $\eta = Y - m(X)$ having mean $0$. But note that (\ref{e:setting-2}) does not capture this most general form since not every $\eta$ can be expressed as $\sigma(X) \epsilon$, with $\epsilon$ independent of $X$.

\begin{remark}
We look at (\ref{e:setting-1}) or (\ref{e:setting-2}) as a ``first-order'' model where interesting relationship between $Y$ and $X$ is captured through the mean function $m(x)$. Other interesting scenarios exist but will not be considered here. For example, $m(x)$ could take one of two different function values $m_1(x)$ and $m_2(x)$, sampled according to some mixture distribution. 
\end{remark}


\section{Methods}
\label{s:methods}


\subsection{Importance sampling scheme and target PDF estimator}
\label{s:mtd-imp}

Recall the notation \eqref{eq:def0}--\eqref{eq:def2} in Section~\ref{s:setting}. Motivated by the discussion in that section, we suggest to take the proposal PDF $\gx$ in \eqref{eq:def2} as
\begin{equation}\label{e:mtd:g}
    \gx = \left\{
    \begin{array}{cl}
    \displaystyle c_L\frac{\fx}{\PL},  & \text{if } \sL ,\\
    \displaystyle c_{0}~\px, & \text{if } \sM,\\
    \displaystyle c_R\frac{\fx}{\PR}, & \text{if } \sR,
    \end{array}
    \right. = 
    \left\{
    \begin{array}{cl}
    \displaystyle c_L~f_X(x|X\le x_L),  & \text{if } \sL ,\\
    \displaystyle c_{0}~\px, & \text{if } \sM,\\
    \displaystyle c_R~f_X(x|X\ge x_R), & \text{if } \sR,
    \end{array}
    \right.
\end{equation}
where $0< c_L, c_{0}, c_R <1$ and $c_L + c_{0} + c_R = 1$, and $f_X(x|A)$ denotes the PDF conditioned on event $A$.

Several comments regarding \eqref{e:mtd:g} are in place. The choice of $ c_L, c_{0}, c_R$ ensures that $\gx$ is a PDF, i.e., it is positive and integrates to 1. It also means that when sampling $N$ observations from \eqref{e:mtd:g}, about $Nc_L$ of the observations should come from $\sL$, $Nc_0$ from $\sM$, and $Nc_R$ from $\sR$. The form of $\gx$ for $\sL$ and $\sR$ is motivated by the discussion in Section~\ref{s:setting}:
for example, for $\sL$, it means effectively that all the observations $\Xio$ with $\Xio \le x_L$ can be included in the sample selected according to \eqref{e:mtd:g}. This is desired as motivated in Section~\ref{s:setting}; see also Figure~\ref{fig:star}. Indeed, the presence of $\fx$ in \eqref{e:mtd:g} means that we sample at random as we did with $\Xio.$ We just need to make sure that $x_L$ is chosen so that there will be about $Nc_L$ observations $\Xio$ with $\Xio \le x_L$. As there are about $N_0~\PL$ such $\Xio$ observations, this will be achieved when
\begin{align}\label{e:mtd:Ncl}
    Nc_L = N_0~\PL.
\end{align}
In practical terms, letting $X_{0,1:N_0} \le X_{0,2:N_0}\le \dots \le X_{0,N_0:N_0}$ be the order statistics of $\Xio$, the relation \eqref{e:mtd:Ncl} holds with 
\begin{align}\label{e:mtd:rl}
    x_L = X_{0, r_L:N_0}, ~r_L = N c_L.
\end{align}
Similarly, to include all observations $\Xio$ with $\Xio \ge x_R$ in the importance sample, we need
\begin{align}\label{e:mtd:Ncr}
    Nc_R = N_0~\PR,
\end{align}
and in practical terms,
\begin{align}\label{e:mtd:rR}
    x_R = X_{0, (N_0-r_R+1):N_0}, ~r_R = N c_R.
\end{align}
Though we present $r_L, r_R$ as resulting from $N, c_L, c_R$, one could fix $r_L, r_R$ in practice, which for fixed $N$, would determine $c_L, c_R$. 
We discuss further the choice of $r_L, r_R$ in Section~\ref{s:est} below. The choice of the PDF $\px$ in \eqref{e:mtd:g} is considered in Section~\ref{s:optimal}. How we sample from $\px$ to obtain one of the observations $\Xio$ with $x_L < \Xio < x_R$ is explained in Section~\ref{s:est}.

If $X_1, \dots, X_N$ denote the sample from the proposal PDF $\gx$, e.g.\ that in \eqref{e:mtd:g}, and $Y_1, \dots, Y_N$ are the corresponding values of $Y$, a natural kernel-based estimator of $\fy$ is then
\begin{equation}\label{e:density-ks}
	\widehat f_Y(y) = \frac{1}{N}\sum_{i=1}^N K_h(y - Y_i) w(X_i),
\end{equation}
where $K_h(u) = h^{-1} K(h^{-1}u)$ for a kernel function $K$ and bandwidth $h>0$, and the weight function $w(x)$ is given by 
\begin{equation}\label{e:density-ks-weights}
	w(x) = \frac{f_X(x)}{g_X(x)}.
\end{equation}
For $\gx$ in \eqref{e:mtd:g}, the weight function is
\begin{equation}\label{e:density-ks-weights-split}
	w(x) = \left\{
    \begin{array}{cl}
 \displaystyle  \frac{1}{c_L}\PL, & \text{ if } \sL ,\\
 \displaystyle  \frac{1}{c_0}\frac{\fx}{\px}, & \text{ if } \sM,\\
 \displaystyle \frac{1}{c_R}\PR, & \text{ if } \sR.
    \end{array}
    \right.
\end{equation}
The kernel function $K$ is assumed to integrate to $1$, that is, $\int K(u) du = 1.$ In practice, we work with the Gaussian kernel $K(u) = \phi(u)$, where $\phi$ is the standard normal density function. On several occasions below, we should refer to the localization property of the kernel function $K$, which states that $\int G(z)K_h(y-z)dz \simeq G(y)$ as $h\rightarrow 0$, for a function $G(z)$ continuous at $z=y$ (\cite{Ghosh2018}). For example, this implies that $\bbE w(X)K_h(y-Y) = \int w(x)f(x,z)K_h(y-z)dx dz \simeq \int w(x)f(x,y)dx$, where $f(x,y)$ is the joint PDF of $(X,Y)$. 

The importance sampling weight function \eqref{e:density-ks-weights} involves the PDF $\fx$ on $\sM$ and the exceedance probabilities $\PL$ and $\PR$. The other quantities ($c_L$, $c_0$, $c_R$, $\px$, $x_L$, $x_R$) are chosen by the user. As noted around \eqref{eq:xlim2}, we assume effectively that $\fx,~\sM$ is estimated well enough to be assumed as known. We shall assume the same about $\PL$ and $\PR$. These issues are considered further in Section~\ref{s:est-t-k}.

Finally, we note that the statistical uncertainty of $\fyhat$ in \eqref{e:density-ks} can be characterized in a straightforward way through
\begin{equation}\label{e:density-ks-var}
	\mbox{\rm Var}(\widehat f_Y(y)) = \frac{1}{N} \mbox{\rm Var}( K_h(y - Y) w(X))
	= \frac{1}{N} \E ( K_h(y - Y)^2 w(X)^2) - \frac{1}{N} (\E K_h(y - Y) w(X))^2.
\end{equation}
The quantity in the parentheses of the second term in (\ref{e:density-ks-var}) can be estimated through (\ref{e:density-ks}). Similarly, for the first term, the expected value can be estimated by $(1/N)\sum_{i=1}^N K_h(y - Y_i)^2 w(X_i)^2$.

\subsection{Optimality of proposal PDF}
\label{s:optimal}

We are interested here in the selection of the proposal PDF $\px$, $\sM$, in \eqref{e:mtd:g} and \eqref{e:density-ks}--\eqref{e:density-ks-weights}. 
By considering the homoscedastic case \eqref{e:setting-1} without the noise $\epsilon$ in Section~\ref{s:opt-nlhm}, we propose the notion of optimality for this selection.
This choice is then examined for the homoscedastic case with noise in Section~\ref{s:opt-hm} and the heteroscedastic case in Section~\ref{s:opt-htr}.

\subsubsection{Noiseless homoscedastic case}
\label{s:opt-nlhm}

We consider here the case \eqref{e:setting-1} with $\sigma=0$, that is,
\begin{equation}\label{e:setting-0}
	Y = m(X).
\end{equation}
We ask what an optimal $\px$ would be in this hypothetical scenario (see also Remark~\ref{rmk:opt2} below) in terms of the variablility of $\fyhat$ in \eqref{e:var-fY}. We consider below two cases: monotone $m$ and piecewise monotone $m$. We assume implicitly that $m$ is differentiable where it is monotone.

\smallskip

\textbf{Monotone ${m}$:}
Consider the case of monotone increasing and differentiable $m$ in \eqref{e:setting-0}.
We have
\begin{equation}\label{e:setting-0-fY-fX}
	f_Y(y) = \frac{f_X(m^{-1}(y))}{m'(m^{-1}(y))}.
\end{equation}
The analogous expression relates $g_Y$ and $g_X$. For these PDFs, recall the definition in \eqref{eq:def0} and \eqref{eq:def2}. Observe further as in (\ref{e:density-ks-var}) that
\begin{equation}\label{e:var-fY}
	N\mbox{Var}(\widehat f_Y(y)) = \mbox{Var}\Big(K_h(y-Y)w(X)\Big) = \E K_h(y-Y)^2 w(X)^2 - (\E K_h(y-Y) w(X))^2. 
\end{equation}
For the second term in (\ref{e:var-fY}), by using the localization property of the kernel function discussed in Section~\ref{s:mtd-imp}, as $h\to 0$,
\begin{equation*}
    \begin{split}
        \E K_h(y-Y) w(X) \ &= \ \E K_h(y-Y) w(m^{-1}(Y)) \\
        &\simeq \ g_Y(y) w(m^{-1}(y)) = g_Y(y)  \frac{f_X(m^{-1}(y))}{g_X(m^{-1}(y))} = \ \frac{f_X(m^{-1}(y))}{m'(m^{-1}(y))}
= f_Y(y),
    \end{split}
\end{equation*}
where we used (\ref{e:setting-0-fY-fX}) twice, with $g_Y,g_X$ first and then with $f_Y,f_X$. Similarly, for the first term in (\ref{e:var-fY}), by setting $\|K\|_2^2 = \int_\bbR K(u)^2 du$ and considering the kernel function $K_2(u) = K(u)^2/\|K\|_2^2$, 
\begin{equation}\label{e:var-fY-firstterm}
\begin{split}
\E K_h(y-Y)^2 w(X)^2 
 &=  \frac{\|K\|_2^2}{h} \E K_{2,h}(y-Y) w(m^{-1}(Y))^2 \\
 &\simeq \frac{\|K\|_2^2}{h} g_Y(y) w(m^{-1}(y))^2 \\
    &= \frac{\|K\|_2^2}{h} g_Y(y)  \frac{f_X(m^{-1}(y))^2}{g_X(m^{-1}(y))^2} \\
&= \frac{\|K\|_2^2}{h}  \frac{f_X(m^{-1}(y))^2}{g_X(m^{-1}(y)) m'(m^{-1}(y))} \\
& = \frac{\|K\|_2^2}{h} f_Y(y)^2 \frac{m'(m^{-1}(y))}{g_X(m^{-1}(y))}.
\end{split}
\end{equation}
Thus, with sufficiently small $h$, the following approximation can be derived:
\begin{equation}\label{e:setting-0-var}
	\frac{ N \mbox{Var}(\widehat f_Y(y)) }{f_Y(y)^2} \simeq \frac{\|K\|_2^2}{h} \frac{m'(m^{-1}(y))}{g_X(m^{-1}(y))} - 1.
\end{equation}
When $g_X$ is set to be the PDF $f_X$, this becomes
\begin{align}\label{e:var-actual}
\frac{ N \mbox{Var}(\widehat f_Y(y)) }{f_Y(y)^2} \simeq \frac{\|K\|_2^2}{h} \frac{1}{f_Y(y)} - 1,
\end{align}
by using (\ref{e:setting-0-fY-fX}). That is, the normalized variance will typically be larger in the distribution tails, where $f_Y(y)$ is smaller. (A separate but related issue is whether one has data of $Y$ in the tails in the first place; this issue should be kept in mind in subsequent developments.)

The notion of optimality that we adopt is to require that the variance of the estimator $\widehat f_Y(y)$, relative to $f_Y(y)$, is constant across $y$.
As optimality concerns the propoal PDF $\px$ defined for $\sM$, we consider $m(x_L)<y<m(x_R)$. That is, we seek:
\begin{equation}\label{e:setting-0-optim}
	\mbox{\bf Optimality} :\frac{ N \mbox{Var}(\widehat f_Y(y)) }{f_Y(y)^2} \simeq \mbox{const}, \quad m(x_L)<y<m(x_R).
\end{equation}
In view of (\ref{e:setting-0-var}), the optimality translates into ${m'(m^{-1}(y))}/{g_X(m^{-1}(y))}$ being constant or 
\begin{equation}\label{e:setting-0-optim-g}
	g_X(x) \propto m'(x), \quad \sM.
\end{equation}
Since $\gx\propto \px$ for $\sM$, this translates into:
\begin{equation}\label{e:setting-0-optim-p}
	\mbox{\bf Optimal proposal PDF} : p_X(x) = C m'(x) = \frac{m'(x)}{m(x_R) - m(x_L)},\quad \sM.
\end{equation}

\begin{example}\label{ex:setting-0-linear}
If $m(x) = ax$, the optimal $p_X$ in (\ref{e:setting-0-optim-p}) is $\px = 1/(x_R-x_L)$, $x\in (x_L,x_R)$, that is, it is uniform on $(x_L,x_R)$. {We remark that uniform sampling is optimal in our criteria for linear relationships.}
\end{example}
\begin{remark}\label{rmk:opt2}
We emphasize again that the setting \eqref{e:setting-0} is hypothetical, serving as a means to investigate what an optimal choice of $p_X$ could be therein. The suggested optimal choice of $p_X$ is investigated in more realistic scenarios in Sections~\ref{s:opt-hm} and \ref{s:opt-htr}. We note that \eqref{e:setting-0} is trivial as far as the main goal of estimating $f_Y$ goes since \eqref{e:setting-0-fY-fX} provides an exact relation to get it from $f_X$ (which we assume to be known for $\sM$).
\end{remark}

\begin{remark}
\label{rmk:opt-def}
Assuming $g_X(x)=p_X(x)$ for simplicity, the proposed optimal PDF in \eqref{e:setting-0-optim-p} ensures the constant relative variance in \eqref{e:setting-0-var} as:
for $m(x_L) < y < m(x_R)$,
\begin{align}
\label{e:seggint-0-var-rmk}
 \frac{ \mbox{Var}(\widehat f_Y(y)) }{f_Y(y)^2} \simeq \frac{\|K\|_2^2 \left(m(x_R)-m(x_L)\right)}{Nh}  - \frac{1}{N}.   
\end{align}
Note that there is no guarantee a priori that the constant on the right-hand side of \eqref{e:seggint-0-var-rmk} is small. But the constant can be made as small as desired by choosing large enough $N$, that is, sampling sufficiently many data points. 
\end{remark}

\begin{remark}
\label{rmk:opt-equiv}
Another interpretation of the optimality criterion \eqref{e:setting-0-optim-p} is to consider the integrated (relative) error, namely, in view of \eqref{e:setting-0-var}, the quantity 
\begin{align}
\label{e:rmk-mise}
    \int_{m(x_L)}^{m(x_R)} \frac{m'(m^{-1}(y))}{g_X(m^{-1}(y))} dy = \int_{x_L}^{x_R} \frac{(m'(x))^2}{g_X(x)}dx.
\end{align}
What density $g_X$ minimizes \eqref{e:rmk-mise}? If one is willing to assume smoothness of $g_X$, the Euler-Lagrange equation (with the Lagrange multiplier for the density constraint) leads to $g_X$ satisfying:
for some constant $C$,
\begin{align}
-\frac{(m'(x))^2}{g_X(x)^2} + C = 0,
\end{align}
that is, $g_X(x)\propto m'(x)$ as in our optimality criterion \eqref{e:setting-0-optim-g}. While considering \eqref{e:rmk-mise} leads to the same optimality criterion, there may be other interesting criteria to consider based on different objectives.
\end{remark}

\textbf{Piecewise monotone ${m}$:}
The arguments above extend easily to the case of piecewise monotone $m$. For such $m$, we partition $(x_L,x_R)$ into intervals $\{A_j\}_{j=1}^n$ so that, when $m$ is restricted to $A_j$, the resulting function $m_j: A_j \mapsto \bbR$ is monotone.
Let $\ylyr$ be the (interior) range of $m$ on $(x_L,x_R)$.
For the developments below, we need to assume that the values of $m(x)$ are outside the range $\ylyr$ where $x\le x_L$ or $x\ge x_R$. When larger values of $y$ are expected for larger values of $x$, this effectively assume that $m(x) \ge m(x_R)$ for $x\ge x_R$ and similarly for $x\le x_L$.
Under the assumptions above, note that 
\begin{equation}
\label{e:PM:fY}
f_Y(y) = \sum\limits_{j=1}^n \frac{f_X(m_j^{-1}(y))}{|m'(m_j^{-1}(y))|}\bbone(y \in m(A_j)), \quad \yM,
\end{equation}
where $\bbone(y\in B)$ is the indicator function for a set $B$. The analogous relation holds for $g_Y$ and $g_X$ replacing $f_Y$ and $f_X$. Then, by arguing as in \eqref{e:var-fY-firstterm} above, one has that
\begin{equation}\label{e:var-fY-firstterm-gen}
\E K_h(y-Y)^2 w(X)^2 \simeq \frac{\|K\|_2^2}{h} \sum\limits_{j=1}^n \frac{f_X(m_j^{-1}(y))^2}{g_X(m_j^{-1}(y)) |m'(m_j^{-1}(y))|}\bbone(y \in m(A_j)), \quad \yM.
\end{equation}
When will this be proportional to $f_Y(y)^2$? In view of \eqref{e:PM:fY}, one can achieve the desired optimality relationship (\ref{e:setting-0-optim}) by requiring
\begin{equation*}
g_X(m_j^{-1}(y)) \propto \frac{f_X(m_j^{-1}(y))}{f_Y(y)}, \quad y \in m(A_j), ~j=1,\dots,n,
\end{equation*}
or, equivalently,
\begin{equation}\label{e:setting-0-optim-g2}
g_X(x) \propto \frac{f_X(x)}{f_Y(m(x))}, \quad \sM.
\end{equation}
In the monotone case $n=1$, $f_Y(y) = f_X(x)/|m'(x)|$ and we find that $g_X(x) \propto |m'(x)|$ as in (\ref{e:setting-0-optim-g}).
Since $\gx \propto \px, \sM$, we propose to require:
\begin{equation}\label{e:setting-0-optim-p2}
\mbox{\bf Optimal proposal PDF} : p_X(x) \propto \frac{f_X(x)}{f_Y(m(x))}, \quad \sM,
\end{equation}
where $f_Y$ is given by \eqref{e:PM:fY}.

\begin{example}\label{ex:setting-0-quadratic}
Let $m(x) = x^2$, $x_L=-1, x_R=1$ and $f_X(x)=c,~x\in (-1,1)$ (i.e. $X$ is uniformly distributed on $(-1,1)$ when conditioned to this interval). $m$ is clearly not monotone but is monotone on the intervals $A_1=(-1, 0]$ and $A_2=(0, 1)$. The range of $m$ over $(-1,1)$ is $[0,1)$ and hence $(y_L,y_R)=(0,1)$. Let $m_1$ and $m_2$ be the functions obtained by restricting $m$ to these intervals, $A_1$ and $A_2$ respectively. For any $y \in (0,1)$, $|m'(m_1^{-1}(y))| = |m'(m_2^{-1}(y))| = 2 \sqrt{y}$. It follows that
\begin{equation*}
f_Y(y) = \frac{f_X(m_1^{-1}(y))}{|m'(m_1^{-1}(y))|} + \frac{f_X(m_2^{-1}(y))}{|m'(m_2^{-1}(y))|} = \frac{c}{2\sqrt{y}} + \frac{c}{2\sqrt{y}} = \frac{c}{\sqrt{y}}, \quad 0<y<1.
\end{equation*}
Thus, the relation \eqref{e:setting-0-optim-p2} becomes
\begin{equation*}
p_X(x) \propto \frac{f_X(x)}{f_Y(m(x))} = \frac{c}{c/|x|} = |x|, \quad -1 < x < 1.
\end{equation*}
This has the behavior we expect: the optimal proposal prefers points closer to the boundary of the support of $X$, which are lower probability points for $Y$. See also the remark below.
\end{example}

\begin{example}\label{ex-setting-0-m2}
Let $x_L=-3, x_R=3, f_X\sim \calN(0,1)$ and 
$$m(x) = \left\{
    \begin{array}{cl}
   18(x+1.2) + 12  & \text{ if } x \le -1.2 ,\\
   -10 x, & \text{ if } -1.2 < x\le 1.2,\\
   18(x-1.2) - 12, & \text{ if } x>1.2.
    \end{array}
    \right.$$
Here, the monotone function $m_1, m_2, m_3$ are linear and defined on intervals $A_1=(-3,-1.2], A_2=(-1.2,1.2], A_3=(1.2,3)$ respectively.
For every value in the range $-12<y<12$, there is $x_i$ such that \( x_i=m_i^{-1}(y) \in A_i \) for \( i=1,2,3 \). Given the relation \eqref{e:setting-0-optim-p2} and the fact that \( f_X \) is standard normal, among \( x_1, x_2, \) and \( x_3 \), the point \( x_2 \) will have a higher value in the proposal PDF since the denominators are the same and \( x_2 \) is closer to the peak of \( f_X \).
This example is further explored in Section~\ref{s:data}. Observing the panel labeled ``m2, known m" in Figure~\ref{f:ns:mean}, it is evident that for \( -12<y<12 \), most $x$-values are sampled in the central linear region.
Note that the weight assigned to the obtained sample will be as in \eqref{e:density-ks-weights-split} and, when combined with \eqref{e:setting-0-optim-p2}, it results in \( w(x) \propto f_Y(m(x)) \). Thus, regardless if sample values come from \( A_1, A_2, \) or \( A_3 \), their contributions to \( \widehat f_Y \) in \eqref{e:density-ks} will be the same. Figure~\ref{f:ns:f1m2} confirms that the target density estimation using the optimal proposal PDF performs well.
\end{example}

\begin{remark}\label{rmk:opt1}
For monotone $m$, the suggested form of the proposal PDF $\px$ is given by \eqref{e:setting-0-optim-p}.
This suggests that the favored regions for sampling are determined by the rate of change of $Y$ with respect to $X$. 
That is, if the change in $Y$ with respect to $X$ is slight, a small sample from that region would be sufficient to estimate the distribution of $Y$. Conversely, if the change in $Y$ in relation to $X$ is abrupt, a higher sampling rate is necessary to accurately estimate the distribution of $Y$.
\end{remark}

\subsubsection{Homoscedastic case}
\label{s:opt-hm}

In the case (\ref{e:setting-1}) with $\sigma>0$, many of the arguments above could be repeated but the resulting expressions do not allow for a closed form solution as in (\ref{e:setting-0-optim-g}). We shall indicate instead what the optimal choice (\ref{e:setting-0-optim-g}) entails in the case (\ref{e:setting-1}) when $\sigma>0$. Assume first monotone increasing $m$. Let 
$$
\widetilde Y = m(X)
$$
so that $Y = \widetilde Y + \sigma \epsilon$, and 
$$
f_Y(y) = \int f_{\widetilde Y}(y-z) \frac{1}{\sigma} f_\epsilon(\frac{z}{\sigma}) dz,
$$
where $f_{\widetilde Y}$ and $f_\epsilon$ are the PDFs of $\widetilde Y$ and $\epsilon$, respectively.

For the second term in the variance (\ref{e:var-fY}), we have 
$$
\E K_h(y-Y) w(X) \simeq \int w(x) f(x,y) dx = \int w(x) f(y|x) g_X(x) dx = \int f(y|x) f_X(x) dx = f_Y(y), 
$$
where $f(x,y)$ and $f(y|x)$ refer to the joint and conditional PDFs, respectively. For the first term in the variance (\ref{e:var-fY}), arguing similarly as in the noiseless case (the asymptotic relation $\simeq$ in (\ref{e:var-fY-firstterm})), we have 
$$
\E K_h(y-Y)^2 w(X)^2 = \frac{\|K\|_2^2}{h} \E K_{2,h}(y-Y) w(X)^2 = \frac{\|K\|_2^2}{h} \E \Big( \E( K_{2,h}(y-\sigma\epsilon - \widetilde Y) w(X)^2 | \epsilon) \Big) 
$$
$$
\simeq \frac{\|K\|_2^2}{h} \E \Big( g_{\widetilde Y}(y-\sigma\epsilon) w(m^{-1}(y-\sigma\epsilon))^2 \Big) = \frac{\|K\|_2^2}{h} \E \Big( f_{\widetilde Y}(y-\sigma\epsilon)^2 \frac{m'(m^{-1}(y-\sigma\epsilon))}{g_X(m^{-1}(y-\sigma\epsilon))} \Big). 
$$
If the optimal choice (\ref{e:setting-0-optim-g}) is used, this becomes
$$
\E K_h(y-Y)^2 w(X)^2 \simeq C \frac{\|K\|_2^2}{h} \E f_{\widetilde Y}(y-\sigma \epsilon)^2
$$
and hence
\begin{equation}\label{e:setting-1-var}
	\frac{ N \mbox{Var}(\widehat f_Y(y)) }{f_Y(y)^2} \simeq C \frac{\|K\|_2^2}{h}\frac{\E f_{\widetilde Y}(y-\sigma\epsilon)^2}{f_Y(y)^2} - 1.
\end{equation}
Note that
\begin{align}\label{e:setting-1-fYtilde-fY}
\frac{\E f_{\widetilde Y}(y-\sigma\epsilon)^2}{f_Y(y)^2} = \frac{ \int f_{\widetilde Y}(y-z)^2 \frac{1}{\sigma} f_\epsilon(\frac{z}{\sigma}) dz}{ (\int f_{\widetilde Y}(y-z) \frac{1}{\sigma} f_\epsilon(\frac{z}{\sigma}) dz)^2}
\end{align}
describes quantitatively the deviation of (\ref{e:setting-1-var}) from being constant over $y$. The smaller $\sigma$ is, the smaller this deviation is.

The formula (\ref{e:setting-1-var}) generalizes easily to the case of piecewise monotone $m$ over a partition $\{A_i\}_{i=1}^n$ of $(x_L,x_R)$, as considered in connection to (\ref{e:setting-0-optim-g2}). Indeed, we have as in (\ref{e:var-fY-firstterm-gen}),
\begin{equation*}
\E \left(K_{2,h}(y - \sigma \epsilon - \widetilde Y) w(X)^2 \Big| \epsilon\right) \simeq \sum_{j=1}^n \frac{f_X(m_j^{-1}(y - \sigma \epsilon))^2}{g_X(m_j^{-1}(y - \sigma \epsilon)) |m'(m_j^{-1}(y-\sigma \epsilon))|}\bbone(y-\sigma\epsilon\in m(A_j)).
\end{equation*}
Plugging in $g_X(x) = C^{-1} f_X(x)/f_{\widetilde Y}(m(x))$ from (\ref{e:setting-0-optim-g2}) leads to 
\begin{align*}
\E K_h(y - Y)^2 w(X)^2 &= \frac{C\|K\|_2^2}{h} \E \sum_{j=1}^n \frac{f_{\widetilde Y}(y - \sigma \epsilon) f_X(m_j^{-1}(y - \sigma \epsilon))^2}{f_X(m_j^{-1}(y - \sigma \epsilon)) |m'(m_j^{-1}(y - \sigma \epsilon))|}\bbone(y-\sigma\epsilon\in m(A_j))\\
&= \frac{C\|K\|_2^2}{h} \E f_{\widetilde Y}(y - \sigma \epsilon) \sum_{j=1}^n \frac{f_X(m_j^{-1}(y - \sigma \epsilon))}{|m'(m_j^{-1}(y - \sigma \epsilon))|}\bbone(y-\sigma\epsilon\in m(A_j)) \\
&= \frac{C \|K\|_2^2}{h} \bbE f_{\widetilde Y}(y - \sigma \epsilon)^2.
\end{align*}
Thus,
\begin{equation}\label{e:setting-1-var-again}
	\frac{ N \mbox{Var}(\widehat f_Y(y)) }{f_Y(y)^2} \simeq \frac{C \|K\|_2^2}{h} \frac{\E f_{\widetilde Y}(y - \sigma \epsilon)^2}{f_Y(y)^2} - 1,
\end{equation}
which agrees with (\ref{e:setting-1-var}) when $m$ is monotone.

\subsubsection{Heteroscedastic case}
\label{s:opt-htr}

We suggest to think of the heteroscedastic case (\ref{e:setting-2}) in more practical terms, namely, as the problem of variance stabilization through a traditional Box-Cox transformation. For example, if $m(x)>0$, $1+\epsilon>0$ and $\sigma(x)=m(x)$, then
\begin{equation}\label{e:setting-2-log}
	\log Y = \log {m(X)} + \log (1+\epsilon) =: \widetilde m(X) + \widetilde \eta
\end{equation}
allows one to fall back to the homoscedastic case (\ref{e:setting-1}). We explore here the implications of such transformations on our problem of interest.

More generally, suppose that
\begin{equation}\label{e:setting-2-BC}
	Z: = \tau_p(Y) = \left\{ 
	\begin{array}{cl}
	\frac{Y^p-1}{p}, & p>0\\
	\log Y, & p=0
	\end{array}
	\right\}
	= \widetilde m(X) + \widetilde \eta,
\end{equation}
where $\tau_p(y)$ is the Box-Cox transformation. (Note that this assumes implicitly that $Y>0$.) The choice $p=0$ is considered in (\ref{e:setting-2-log}) and corresponds to $Y = e^Z= e^{\widetilde m(X)} e^{\widetilde \eta} = e^{\widetilde m(X)} \E e^{\widetilde \eta} + e^{\widetilde m(X)} (e^{\widetilde \eta}- \E e^{\widetilde \eta})$, that is, the heteroscedastic case $m(x) = e^{\widetilde m(X)} \E e^{\widetilde \eta}$ and $\sigma(x)\propto m(x)$. For $p=1/k$, $k\in\bbN$, it follows from (\ref{e:setting-2-BC}) that
$$
Y = (pZ+1)^{1/p} = (p \widetilde m(X) + 1 + p \widetilde \eta)^{1/p} = (\bar m(X) + \bar \eta)^k = \bar m(X)^k + k \bar m(X)^{k-1} \bar \eta + \ldots, 
$$
where $\bar m(x) = p\widetilde m(x) + 1$ and $ \bar \eta = p \widetilde \eta$. This case corresponds approximately to the heteroscedastic case 
\begin{equation}\label{e:setting-2-relations}
	m(x) = \bar m(x)^k = (p\widetilde m(x) + 1)^{1/p},\quad \sigma(x) \propto \bar m(x)^{k-1} = m(x)^{1-1/k} = m(x)^{1-p}.
\end{equation}
E.g., for $p=1/2$, $\sigma(x) \propto m(x)^{1/2}$.

It is interesting to examine the effect of the transformation (\ref{e:setting-2-BC}) on our choice of optimal proposal density (\ref{e:setting-0-optim-p}). Continuing with the above case $p=1/k$, $k\in\bbN$, note that (\ref{e:setting-2-relations}) implies that
$$
\widetilde m(x) = \frac{m(x)^p-1}{p} = \tau_p(m(x))
$$
and that the optimal $g_X$ is
\begin{equation}\label{e:setting-2-optimal}
	p_X(x) \propto \widetilde m'(x) \propto m(x)^{p-1}, \quad \sM. 
\end{equation}

\begin{example}\label{ex:setting-2-sqroot}
For $p=1/2$, $m(x)=x$, $\sigma(x)\propto x^{1/2}$, the choice (\ref{e:setting-2-optimal}) yields $p_X(x) \propto x^{-1/2}$. In contrast, without the transformation in the homogeneous case of this example, $p_X(x)\propto 1$.
\end{example}

Another issue in the heteroscedastic case is what density is exactly estimated ($f_Y$ or $f_Z$), and through what method. The discussion above involves a transformation to go from $Y$ to $Z$, and subsequent optimality refers to estimating $f_Z$ as in (\ref{e:density-ks}), that is, 
\begin{equation}\label{e:setting-2-density-ks}
	\widehat f_Z(z) = \frac{1}{N}\sum_{i=1}^N K_h(z - Z_i) w(X_i).
\end{equation}
As $f_Y(y) = \tau_p'(y) f_Z(\tau_p(y))$, on one hand, it is natural to set
\begin{equation}\label{e:setting-2-densities-YZ}
	\widehat f_Y(y) = \tau_p'(y) \widehat f_Z(\tau_p(y)).
\end{equation}
Note that with this choice,
\begin{equation}\label{e:setting-2-densities-YZ-var}
	\frac{\mbox{Var}(\widehat f_Y(y)) }{f_Y(y)^2} = \frac{\mbox{Var}(\widehat f_Z(\tau_p(y))) }{f_Z(\tau_p(y))^2}.
\end{equation}
So, for example, if the right-hand side of (\ref{e:setting-2-densities-YZ-var}) is (nearly) constant, then so is the left-hand side.

On the other hand, we also note that the estimator (\ref{e:setting-2-densities-YZ}) is close to a kernel-based estimator of $f_Y$ obtained directly from $Y_i$ in the following sense. Indeed,
$$
\frac{1}{N}\sum_{i=1}^N K_h(y - Y_i) w(X_i) = \frac{1}{N}\sum_{i=1}^N K_h\Big(y - \tau_p^{-1}(Z_i)\Big) w(X_i) = \frac{1}{N}\sum_{i=1}^N K_h\Big(\tau_p^{-1}(\tau_p(y)) - \tau_p^{-1}(Z_i)\Big) w(X_i)
$$
$$
\simeq \frac{1}{N}\sum_{i=1}^N K_h\Big((\tau_p^{-1})'(\tau_p(y)) (\tau_p(y) - Z_i)\Big) w(X_i) 
= \tau_p'(y) \frac{1}{N}\sum_{i=1}^N K_{h\tau_p'(y)}(\tau_p(y) - Z_i) w(X_i)
$$
or
$$
\frac{1}{N}\sum_{i=1}^N K_{h/\tau_p'(y)}(y - Y_i) w(X_i) \simeq \tau_p'(y) \widehat f_Z(\tau_p(y)).
$$
That is, one can think of the estimator (\ref{e:setting-2-densities-YZ}) as the kernel-based estimator of $f_Y(y)$ but using a location-dependent bandwidth.

\section{Modified estimator for target PDF tails}
\label{s:gpd}

The estimator $\fyhat$ in \eqref{e:density-ks} is defined for any $y$ in principle. As with the estimation of $f_X$ discussed in Sections~\ref{s:setting} and \ref{s:est-t-k}, however, the estimator $\fyhat$ is expected to be meaningful only for $y\in\ylyr$ and suitable $y_L, y_R$. For example, one could naturally expect $\min_{i=1,\dots,N}{Y_i} \le y_L$ and $y_R \le \max_{i=1,\dots,N}{Y_i}$. We discuss the choice of $y_L, y_R$ in Section~\ref{s:est-t-k} and also in connection to the presentation below, in Section~\ref{s:est-t-gpd}. We consider here a natural way to estimate $\fy$ beyond the thresholds $y_L$ and $y_R$.

The idea is to exploit the so-called second extreme value theorem, or the Pickands–Balkema–De Haan theorem, stating that (essentially) any distribution above high enough threshold can be approximated by the generalized Pareto distribution (GPD). See, for example, \cite{coles:2001} and \cite{Embrechts}. Motivated by this observation, we define our final estimator of the target PDF $\fy$ as
\begin{equation}\label{e:fyhatm}
    \fyhatm = \left\{ 
    \begin{array}{cl}
        \widehat c'_R \cdot g_{\hxi_R, \hdel_R}(y-y_R), & \text{if } \yR,\vspace{1mm}\\
        \fyhat, & \text{if } \yM, \\
        \widehat c'_L \cdot g_{\hxi_L, \hdel_L}(-(y-y_L)), & \text{if } \yL.\\
    \end{array}
    \right.
\end{equation}
Here, $\fyhat$ is given by \eqref{e:density-ks}, $\widehat c'_L$ and $\widehat c'_R$ are normalizing constants, and $\gpd$ is the PDF of GPD given by
\begin{equation}\label{e:gpd}
    \gpd = \left\{ 
    \begin{array}{cl}
        \gfunction,~u>0, & \text{if } \xi>0,\\
   \displaystyle     \frac{1}{\beta}e^{-\frac{u}{\beta}},~u>0, & \text{if } \xi=0, \\
        \gfunction,~0<u<-\frac{\beta}{\xi}, & \text{if } \xi<0,\\
    \end{array}
    \right.
\end{equation}
where $\xi$ and $\beta$ are the shape and scale parameters.
The GPD parameter estimates $\hxi_R, \hdel_R$ in \eqref{e:fyhatm} are based on the data $Y_i > y_R$, and $\hxi_L, \hdel_L$ on the data $Y_i< y_L$. In practice, we use maximum likelihood estimation and, more precisely, its weighted version, since $Y_i$'s are obtained from importance sampling. The importance sampling weights are given by $w(X_i)$ with $w(x)$ defined in \eqref{e:density-ks-weights-split}.
Additionally, we use
\begin{equation}\label{e:gpd:cL}
    \widehat c'_R = \widehat \bbP (Y \ge y_R) =\frac{1}{N}\sum_{i=1}^N \bbone(Y_i \ge y_R)w(X_i),
\end{equation}
and analogously for $\widehat c'_L$ as the normalizing constants. Since $\widehat c'_L,\widehat c'_R$ and $\fyhat$ in \eqref{e:fyhatm} are estimates, the estimator \eqref{e:fyhatm} need not integrate exactly to one. Thus, additional normalization can be applied if needed.
Numerical illustrations are postponed till Section~\ref{s:data}.

\section{Related sampling and estimation issues}
\label{s:est}
We first introduce a sampling algorithm based on the proposal PDF \eqref{e:mtd:g} in Section~\ref{s:est:sampler}. We then discuss the estimation of the mean function in Section~\ref{s:est-m}, followed by the selection of thresholds in Section~\ref{s:est-t}. 

\subsection{Sampling low-fidelity outputs by proposal PDF}
\label{s:est:sampler}

We discuss here how to sample $N$ pairs $(X_i, Y_i)$ based on the distribution $g_X$ proposed in \eqref{e:mtd:g}.
As $X$ represents the less expensive low-fidelity outputs, we first generate the set $\mathcal{X}_0 = \{X_{0,1}, \dots, X_{0,N_0}\}$ through $N_0$ distinct random seeds. This set ${\cal X}_0$ serves two primary purposes: it provides a baseline set of $X$ values upon which further sampling can be applied, and it enables the generation of the corresponding $Y$ values, since each $\Xio$ in ${\cal X}_0$ is linked to a specific random seed that can be used to produce its $Y$ counterpart.

To sample $N$ values from $g_X$, we refer to the discussion in Section~\ref{s:mtd-imp}. Specifically, we expect about $r_L$ sample points in $(-\infty,x_L]$, $r_R$ sample points in $[x_R,\infty)$, and the rest in $(x_L,x_R)$.
Also, we define $x_L$ as the smallest $r_L$th order statistic from $\mathcal{X}_0$ as in \eqref{e:mtd:rl} and similarly for $x_R$ in \eqref{e:mtd:rR}.
Accordingly, we include all $X_{0,i}$ values where $X_{0,i}\le x_L$, and analogously for $X_{0,i}\ge x_R$. For the range $x_L < X < x_R$, we first sample $N-r_L-r_R$ values from $p_X$, and then pick the nearest neighbor from ${\cal X}_0$ without replacement.
As a result, we are able to sample $N$ values of $X$ from ${\cal X}_0$, which allows for the generation of the corresponding $Y$ values via the shared underlying random seed. The procedure is summarized in Algorithm~\ref{a:ISsampler}.

\begin{algorithm}
\begin{algorithmic}[1]
\Statex \textbf{Input:} initial parameters \(N_0\), \(N\), \(r_L\), \(r_R\)
\State sample \(X_{0,1}, \dots, X_{0,N_0}\) from \(f_X\)
\State determine \(x_L=X_{0,r_L:N_0}\), \(x_R=X_{0,(N_0-r_R+1):N_0}\)
\State sample \(N-r_L -r_R\) points from \(p_X\), find the nearest neighbor in \(\{X_{0,1}, \dots, X_{0,N_0}\}\) without replacement, and store these values as \(X_{r_L+1}, \dots, X_{N-r_R}\)
\State obtain \(\{X_1, \dots, X_N\} = \{X_{0,i:N_0},~i\le r_L\}\)\(\cup \{X_i,~r_L+1\le i \le N-r_R\}\)\(\cup \{X_{0, i:N_0},~i\ge N_0-r_L+1\}\)
\State sample \(Y_i\) given \(X=X_i\)
\Statex \textbf{Output:} Sample \((X_1, Y_1), \dots, (X_N, Y_N)\).
\end{algorithmic}
\caption{Sampling Strategy from Proposal PDF \eqref{e:mtd:g}}
\label{a:ISsampler}
\end{algorithm}

\subsection{Estimation of mean function}
\label{s:est-m}

As discussed in Section~\ref{s:optimal}, the proposal PDF $p_X$ is constructed using both the mean function $m$ and the PDF $f_X$. However, since $m$ is not commonly known in practice, our sampling scheme should be adapted to include its estimation.
This section elaborates on how we modify our sampling scheme to progressively learn and update estimates of $m$ and $p_X$.

To begin our sampling scheme, we need to obtain an initial estimate of $m$, denoted as $\widehat m^{(0)}$. To do this, we start by obtaining a small initial set of $(X_{-1}, Y_{-1}), \ldots, (X_{-n_0}, Y_{-n_0})$, where $X$ is sampled uniformly within the range $(x_L, x_R)$. Note that the sampling follows analogously the procedure in Section~\ref{s:est:sampler}. Then, we propose to use piecewise linear regression (PLR) explained in more detail below to derive an estimate for $m$ that best fits this data.
The rest of our sampling scheme is iterative in nature. At iteration $t$, given that we have $\widehat m^{(t-1)}$, we estimate $\widehat p_X$ based on \eqref{e:setting-0-optim-p2} using the plug-in estimator $\widehat m^{(t-1)}$. We then draw a new $X_{t}$ from $ \widehat p_X$ and its corresponding $Y_{t}$. Then, we obtain $\widehat m^{(t)}$ using the updated dataset. This process is repeated until we collect the desired sample of size $\tilde N$, i.e., $t=1, \dots, \tilde N$. In particular, when estimating the target PDF $f_Y$, the initial $n_0$ data points are excluded.

Regarding the specifics of estimating $m$, we employ piecewise linear regression (PLR), which presents several advantages. PLR ensures monotonicity within each segment, allowing for straightforward computations of inverse functions and derivatives.
To be specific, suppose that the resulting monotone components are $\{\widehat{m}_1, \ldots, \widehat{m}_{J}\}$. Each component $\widehat{m}_j$ is linear and defined over the interval $A_j = (\Xin, \Xiin)$. The points $ (\Xin, \Yin)$ serve as breakpoints for these piecewise linear segments, such that $x_{(1)}= x_L,$ $x_{(J+1)}= x_R, $ $\Xin\le \Xiin$, and $\Yin = \widehat m_{j}(\Xin)$ for all $j$. Consequently, for $\Xin \leq x \leq \Xiin$, the equation for $\widehat m(x)$ is given by
\begin{equation}\label{plr-mhat}
\widehat{m}(x) = \left(\frac{\Xiin-x}{\Xiin - \Xin}\right) \Yin + \left(\frac{x - \Xin}{\Xiin - \Xin}\right) \Yiin.
\end{equation}
Then, we can derive the quantities needed for \eqref{e:PM:fY} as
\begin{equation}\label{plr-minv}
\widehat{m}_j^{-1}(y) = \frac{(y-\Yiin)(\Xiin - \Xin)}{\Yiin - \Yin} + \Xiin \quad\mbox{and}\quad \widehat{m}_j'(x) = \frac{\Yiin - \Yin}{\Xiin - \Xin}.
\end{equation}
In our numerical studies in Section~\ref{s:data}, we utilized the R package \texttt{segmented} to obtain PLR (\cite{segmented}). This package allows for PLR fitting by specifying the number of break points. By evaluating the Akaike Information Criterion (AIC) for each model with the varying number of break points, we selected the one with the lowest AIC to determine the optimal number of break points as the best-fitted model.

\begin{algorithm}[t]
\caption{Adaptive Sampling Incorprating $m$ Estimation}
\label{a:adaptive}
\begin{algorithmic}[1]
\Statex \textbf{Input:} PDF $f_X$, thresholds $x_L$ and $x_R$
\State sample $(X_i, Y_i)$ where $X_i \sim \text{Unif}(x_L,x_R)$ for $i=-1, \ldots, -n_0$
\State construct $D^{(0)}=\{(X_i,Y_i),  ~i=-1, \ldots, -n_0\} $
\State fit piecewise linear regression (PLR) to $D^{(0)}$ to obtain the initial estimate $\widehat m^{(0)}$ and its monotone components $\{\widehat{m}_{j,0}, j\in \mathcal{J}^{(0)} \}$ 
\For{$t=1,\dots,\tilde N$}
    \State $\widehat f_{\tilde{Y}}^{(t)}(y) \gets \sum_{j\in \mathcal{J}^{(t-1)}} \frac{f_X(\widehat{m}_{j,t-1}^{-1}(y))}{|{\widehat{m}_{j,t-1}}'({\widehat{m}_{j,t-1}}^{-1}(y))|}\bbone(y\in \widehat m^{(t)}(A_j)) $
    \State $\widehat{p}_X^{(t)}(x) \gets \frac{f_X(x)}{\widehat f_{\tilde{Y}}^{(t)}(\widehat{m}^{(t-1)}(x))}$
    \Comment{construct $\widehat p_X$}
    \State normalize $\widehat p_X^{(t)}$ on $\sM$
    \State sample $(X_t, Y_t)$ where $X_{t} \sim \widehat{p}_X^{(t)}$
    \Comment{sample new point}
    \State $w(X_t) \gets \frac{f_X(X_{t})}{\widehat{p}_X^{(t)}(X_{t})}$
    \Comment{update weights}
    \State update $D^{(t)}=\{(X_{-n_0},Y_{-n_0}), \dots, (X_{-1},Y_{-1}), (X_{1},Y_{1}), \dots, (X_{t},Y_{t})\}$
    \State fit PLR to $D^{(t)}$ to obtain $\widehat m^{(t)}$ and its monotone components $\{\widehat{m}_{j,t}, j\in \mathcal{J}^{(t)} \}$ 
    \Comment{update $\widehat m$}
\EndFor
\Statex \textbf{Output:} Sample \((X_1, Y_1), \dots, (X_{\tilde N}, Y_{\tilde N})\).
\end{algorithmic}
\end{algorithm}

Once an estimate $\widehat{m}$ of $m$ is obtained from the observed sample, we draw additional data points according to $\widehat{p}_X$.
If the estimated $\widehat m$ is monotone, the CDF of our proposal PDF $\widehat p_X$, denoted as $\widehat P_X$, is proportional to $\widehat m$, i.e., $\widehat P_X(x) \propto \widehat m(x)$. Using inverse transform sampling, we can then sample $X$ as $\widehat P_X^{-1}(U)$, where $U\sim \text{Unif}(0,1)$, which simply involves random sampling from uniform distribution. When $\widehat m$ is piecewise monotone, sampling techniques such as Metropolis-Hastings or inverse transform sampling can be used (e.g., \cite{robert_monte_2004}).

\begin{figure}[t]
\centering
\includegraphics[width=.8\linewidth]{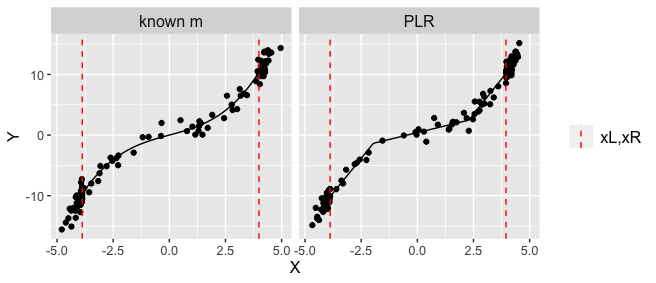}
\caption{Left: Sample obtained from the proposal PDF $p_X$ with known $m$ (Algorithm~\ref{a:ISsampler}) and the true $m$ curve. Right: Sample obtained from the adaptive sampling (Algorithm~\ref{a:adaptive}) and the final fitted PLR curve.}	
\label{f:plr:samples}
\end{figure}

Our sampling procedure incorporating $m$ estimation is summarized in Algorithm~\ref{a:adaptive}. For illustration, Figure~\ref{f:plr:samples} further presents the samples obtained from the proposal PDF $p_X$ using Algorithm~\ref{a:ISsampler} with known $m$ (left) and from the adaptive sampling via Algorithm~\ref{a:adaptive} with unknown $m$ alongside with the final fitted PLR lines (right). The thresholds $x_L$ and $x_R$ are indicated by the red vertical dashed lines in the figure.

\begin{remark}
\label{rmk:est-m-gp}
While this section mainly introduced PLR for function approximation, other methods like Gaussian Process Regression (GPR) and nonparametric kernel regression are also applicable (e.g., \cite{GPR}, \cite{wand1994kernel}). Focusing on GPR, when \( m \) is assumed to have a prior distribution characterized by mean function \( m_0 \) and a positive semi-definite covariance function \( k \), one writes $m \sim \mathcal{GP}(m_0, k)$, implying that for any $n$-dimensional input vector $\mathbf{x} \in \mathbb{R}^n$,
\begin{eqnarray*}
	m(\mathbf{x}) \sim \mathcal{N}(m_0(\mathbf{x}), k(\mathbf{x}, \mathbf{x})).
\end{eqnarray*}
Given independent errors $\sigma\epsilon$ from \eqref{e:setting-1} following a $\mathcal{N}(0, \sigma^2)$ distribution, and observed values $\mathbf{X} = (X_1, \ldots, X_t)$ and $\mathbf{Y} = (Y_1, \ldots, Y_t)$, we have:
\begin{eqnarray*}
	m | \mathbf{X}, \mathbf{Y} \sim \GP(m^*, k^*),
\end{eqnarray*}
where
\begin{eqnarray}
	m^*(x) & = & m_0(x) + k(x, \mathbf{X}) \Big(k(\mathbf{X}, \mathbf{X}) + \sigma^2 \mathbf{I}_t\Big)^{-1} (\mathbf{Y}-m_0(\mathbf{X})), \label{eq:gpr_mean}\\
	k^*(x, \widetilde x) & = & k(x, \widetilde x) - k(x, \mathbf{X}) \Big(k(\mathbf{X}, \mathbf{X}) + \sigma^2 \mathbf{I}_t\Big)^{-1} k(\mathbf{X}, \widetilde x),	\label{eq:gpr_var}
\end{eqnarray}
where $\mathbf{I}_t$ is an $t\times t$ identity matrix.
This conjugacy enables GPR to consistently update the mean function with each new observation. Furthermore, derivative functions are readily accessible without additional computational cost, making GPR an appealing alternative.
\end{remark}

\begin{remark}
\label{rmk:est-m-extreme}
Note that thus far, we treated the distribution of $Y$-values as being equally important over its range. In some applications, however, the interest and relevance might be greater for one of the distribution tails, for example, the right tail associated with larger $Y$-values. We cannot completely decouple the sampling of $X$-values across the two distribution tails as, a priori, larger $Y$-values could also result from smaller $X$-values, especially in the case when the correlation between $X$ and $Y$ is not strong. But if there is indication for strong correlation (as after the preliminary step of Algorithm~\ref{a:adaptive} having $n_0$ data points $(X_i, Y_i)$), note that our subsequent sampling procedure could accommodate the situation where the emphasis is placed on one tail only. For example, the left tail could be de-emphasized by choosing $c_L < c_R$ in \eqref{e:mtd:g}, and we could similarly reweigh the resulting density $\px$, so that more weight is put on larger $X$-values.
\end{remark}

\subsection{Selection of thresholds}
\label{s:est-t}
\subsubsection{Range for kernel-based estimation of PDF}
\label{s:est-t-k}

We assumed in Section~\ref{s:mtd-imp} that the thresholds $x_L$, $x_R$ are given defining the range $\xlxr$ where the PDF $f_X(x)$ can be estimated well, say through the kernel-based estimator
\begin{align}
    \label{e:est:t:fX}
    \widehat f_X(x) = \frac{1}{N_0}\sum_{i=1}^{N_0}K_h(x-\Xio).
\end{align}
Furthermore, as in \eqref{e:mtd:rl} and \eqref{e:mtd:rR}, we formulated the threshold selection as that of $r_L$ and $r_R$ in the order statistics as $x_L=X_{0,r_L:N_0}$ and $x_R=X_{0,(N_0-r_R+1):N_0}$. In this section, we ask what $r_L$ and $r_R$ (or, $x_L$ and $x_R$) should be taken in practice. Put differently, for example in connection to $r_R$, up to what largest value of $\Xio$, could one expect that $\fxhat $ estimates $\fx$ well?

Note that the same question is also relevant for the weighted kernel-based density estimator $\fyhat$ in \eqref{e:density-ks} in view of the modified estimator $\fyhatm$ in \eqref{e:fyhatm} and the selection of thresholds $y_L, y_R$. Furthermore, the choice of $y_L, y_R$ here is connected not only to the range $\ylyr$ for the estimation of $\fyhat$ but also to the use of GPD beyond the two thresholds. The latter issue is discussed in Section~\ref{s:est-t-gpd} below. There is though also a difference in the role played by $\fxhat$ and $\fyhat$ in our approach: while we seek $\xlxr$ where $\fxhat$ can effectively replace $\fx$, this is not quite the goal with $\fyhat$ and $\fyhatm$ which are viewed as estimators with certain uncertainty properties. For this reason, we will focus on the question raised for $\fxhat$ and then make some comments concerning $\fyhat$.

The question above concerning $\fxhat$ seems rather basic but we are not aware of previous works addressing it directly. Addressing it here fully goes beyond the scope of this study. In fact, we shall restrict our discussion to making a few related points and more practical recommendations. We shall consider a related but slightly simpler question, for example concerning the right tail of the distribution, on how large $x_R$ (or $r_R$) one can take so that the empirical tail probability
\begin{equation}
    \label{e:est:t:Fbar}
    \FbhR = \frac{1}{N_0}\sum_{i=1}^{N_0} \bbone(\Xio > x_R)
\end{equation}
estimates the true tail probability $\bar F(x_R) = \bbP(X>x_R)$ well. The first discussion below can be adapted for $\fxhat$ but we are not aware if this has been done for $\fxhat$ with the second discussion below.

First, the question above about $\FbhR$ can be addressed through the following more informal argument. Note that the variance of the estimator is given by
\begin{align}\label{e:est:t:VarFx}
    \var(\FbhR)= \frac{1}{N_0}\Prr(1-\Prr).
\end{align}
Then, the  variance relative to the tail probability is approximately in the tail:
\begin{align}\label{e:est:t:m:rel_var}
    \frac{\var(\FbhR)}{\bar F_X(x_R)^2} = \frac{1}{N_0}\frac{1-\Prr}{\Prr} \simeq \frac{1}{N_0~\Prr} \simeq \frac{1}{r_R},
\end{align}
where $r_R$ is the number of observations $\Xio > x_R$. This suggests that the relative variance could be made small practically speaking when $r_R = 10$ or larger. In our numerical studies in Section~\ref{s:data:ex} and \ref{s:data:other}, we use $r_R=25$.

Second, the informal argument above can be put on a more solid footing as follows. We can similarly seek to understand the behavior of
\begin{align}\label{e:est:Fb-Fbh}
    \frac{\bar F_X(\xRest)}{\widehat{\bar F}_X(\xRest)} = \frac{N_0}{r-1} \bar F_X(\xRest).
\end{align}
As $F_X(X)$ is a uniform random variable $U$ on $(0,1)$, note that $\bar F_X(\xRest)$ is the order statistic $U_{r:N_0}$. It is known (e.g., \cite{arnold2008}) that 
\begin{align}
    \label{e:est:Ur}
    U_{r:N_0} \sim {Beta}(r, N_0+1-r),
\end{align}
where $Beta(\cdot,\cdot)$ denotes the Beta distribution. It follows that 
\begin{align}\label{e:est:Fb-Fbh-beta}
    \frac{\bar F_X(\xRest)}{\widehat{\bar F}_X(\xRest)} \sim \frac{N_0}{r-1} Beta(r,N_0+1-r) =: \xi_{N_0,r}.
\end{align}
Observe that
\begin{align}\label{e:est:Ebeta}
    \bbE\xiNr = \frac{N_0}{r-1}\frac{r}{N_0+1}, \quad \var(\xiNr) = \left(\frac{N_0}{r-1}\right)^2\frac{r(N_0+1-r)}{(N_0+1)^2(N_0+2)}.
\end{align}
As $r$ is increasing, $\var(\xiNr)$ is decreasing and $\bbE\xiNr$ approaches 1 (for large $N_0$), showing that the ratio in \eqref{e:est:Fb-Fbh} will tend to be closer to 1 as well. Furthermore, for any $r$, the ratio in \eqref{e:est:Fb-Fbh} has bounded variability.

We have explored similar questions numerically for the weighted kernel-based estimator $\fyhat$ in \eqref{e:density-ks}. We similarly found that taking, for example, $r$th largest value for the upper bound $y_R$ of the estimation range seemed to control variability, though a deeper study would also be warranted. 

\subsubsection{Generalized Pareto fit}
\label{s:est-t-gpd}

Note that for the modified estimator in \eqref{e:fyhatm}, for example, $y_R$ is not only the upper bound up to which to use $\fyhat$, but also the threshold above which to fit the GPD. From the latter perspective, the threshold selection is a well-studied problem in extreme value theory. The methods range from more ad hoc (e.g., \cite{coles:2001}, Section 4.3.1) to more sophisticated (e.g. \cite{dupuis2006}). They are not the focus of this study. In our numerical studies, we use a fixed number of observation above threshold across different replications.


\section{Numerical studies}
\label{s:data}

This section presents a simulation study and an application to evaluate the performance of the proposed methods. Section~\ref{s:data:ex} contains results for some representative cases, followed by further discussion in Section~\ref{s:data:other} on several related points. The reproducible R code for the presented
simulations is available at \href{https://github.com/mjkim1001/MFsampling}{https://github.com/mjkim1001/MFsampling}. Section~\ref{s:data:ship} contains an application to ship motions.

\subsection{Illustrations for several informative cases}
\label{s:data:ex}
\begin{figure}
\centering
\begin{subfigure}{.9\linewidth}
    \centering
    \includegraphics[width=1\linewidth]{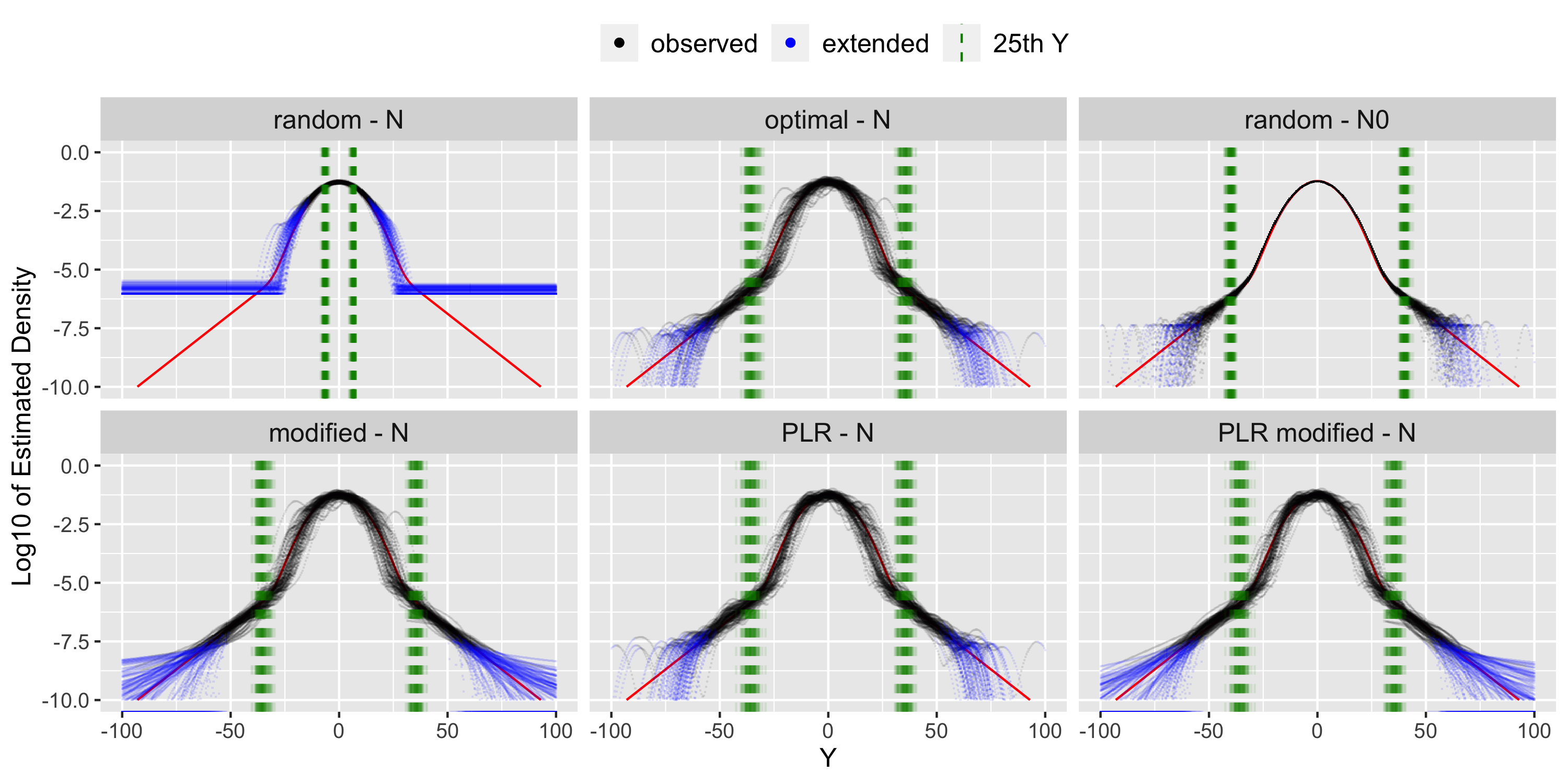}
    \vspace{-2em}
    \caption{Monotone scenario ($m_1$)}
    \label{f:ns:f1m1}
    \vspace{.5em}
\end{subfigure}
\\ 
\begin{subfigure}{.9\linewidth}
    \centering
    \includegraphics[width=1\linewidth]{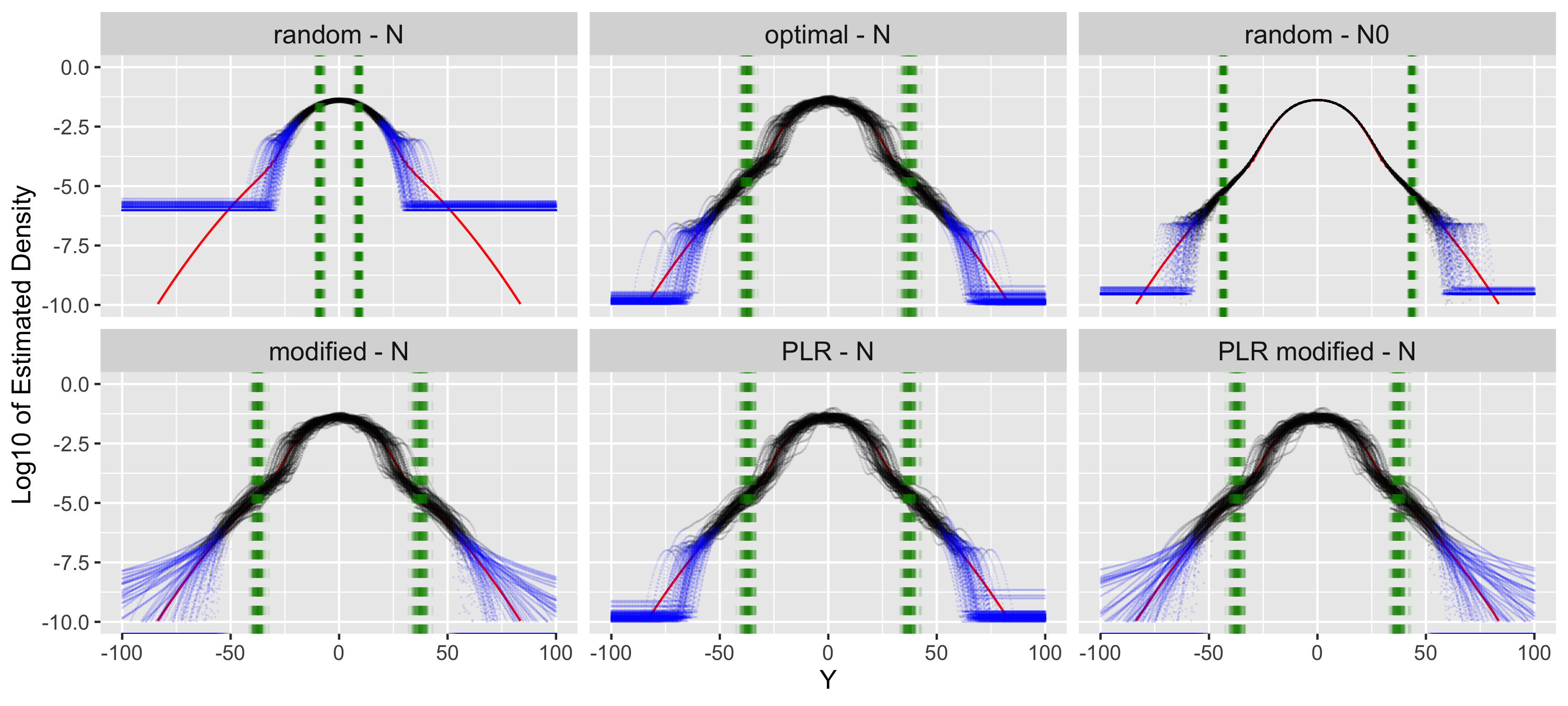}
     \vspace{-2em}
    \caption{Piecewise monotone scenario ($m_2$)}
    \label{f:ns:f1m2}
    \vspace{.5em}
\end{subfigure}
\\
\begin{subfigure}{.9\linewidth}
    \centering
    \includegraphics[width=1\linewidth]{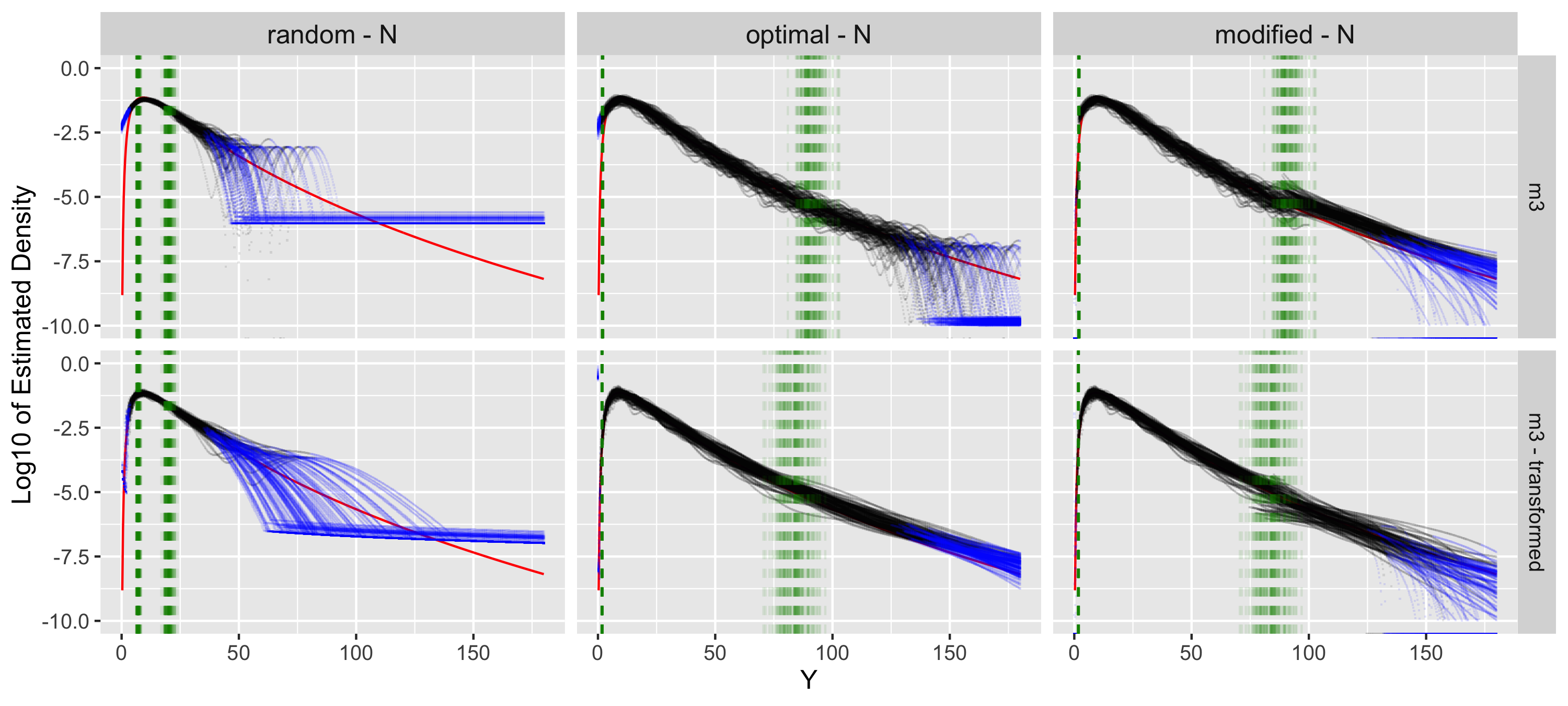}
     \vspace{-2.2em}
    \caption{Heteroscedastic scenario ($m_3$)}
    \label{f:ns:f1m3}
\end{subfigure}
\caption{Estimated versus true log-PDF over 100 realizations for various sampling strategies.}
\label{f:ns:f1}
\end{figure}

In this section, we present numerical illustrations of the proposed density estimators and sampling schemes through a set of informative cases. We present mean functions $m_i$ for three distinct scenarios, $i=1,2,3,$ each representing a different type of relationship:
$m_1$ corresponds to a monotone relation,
$m_2$ to a piecewise monotone relation, and $m_3$, which involves an exponential function, is used to exemplify a heteroscedastic relation.
The mean functions are as follows:
\begin{equation*}\label{e:ns:mean}
\begin{split}
m_1(x) & = 3x,\\
m_2(x) & = \left\{
    \begin{array}{cl}
   18(x+1.2) + 12  & \text{ if } x \le -1.2 ,\\
   -10 x, & \text{ if } -1.2 < x\le 1.2,\\
   18(x-1.2) - 12, & \text{ if } x>1.2,
    \end{array}
    \right.\\
m_3(x) & = e^{x/2}.
\end{split}
\end{equation*}
For examining the piecewise monotone and heteroscedastic scenarios, we generate $X$ from a certain normal distribution. On the other hand, to assess the monotone scenario, we generate $X$ according to the density
$$
f_X^h(x) =  \left\{
    \begin{array}{cl}
    C e^{\frac{1}{2}x-6}, & \text{ if } x \le -4 , \\
   C e^{-\frac{1}{2}x^2},  & \text{ if } -4 < x \le 4 ,\\
   C e^{-\frac{1}{2}x-6}, & \text{ if } x > 4 ,
    \end{array}
    \right.\\
$$
where $C$ is a normalizing constant.
This distribution is constructed to follow a normal distribution at the center and to have heavier tails at the extremes. The distributions of ship motions tend to have such shape (e.g. \cite{BELENKY2019}). Another rationale behind this design is to induce curvature changes at the distribution tails, as can be seen in Figure~\ref{f:ns:f1m1}. This allows investigating how well each estimator captures these variations at the tails and understanding the role of GPD thresholds.

\begin{figure}[ht]
\centering
\begin{subfigure}{.49\linewidth}
    \centering
    \includegraphics[width=1\linewidth]{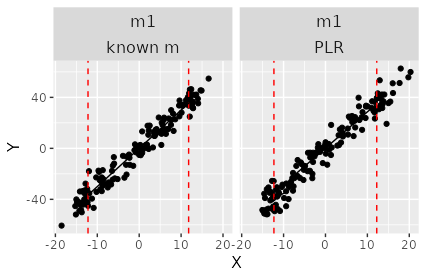}
\end{subfigure}
\begin{subfigure}{.49\linewidth}
    \centering
    \includegraphics[width=1\linewidth]{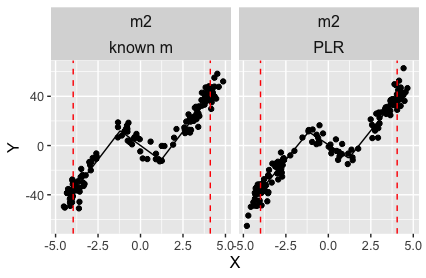}
\end{subfigure}
\\ 
\begin{subfigure}{.49\linewidth}
    \centering
    \includegraphics[width=1\linewidth]{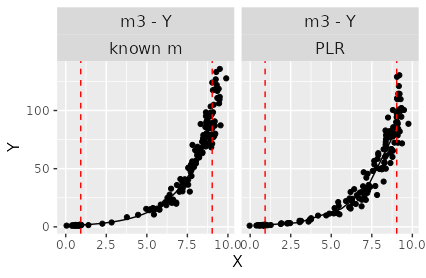}
\end{subfigure}
\begin{subfigure}{.49\linewidth}
    \centering
    \includegraphics[width=1\linewidth]{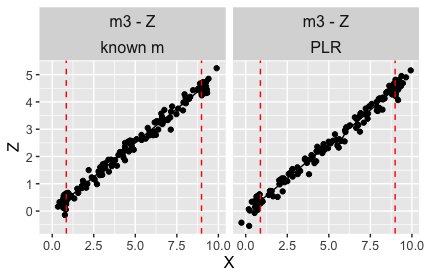}
\end{subfigure}
\caption{Mean functions $m$, their PLR estimates, and the samples obtained via Algorithms~\ref{a:ISsampler} (known $m$) and \ref{a:adaptive} (PLR).}
\label{f:ns:mean}
\end{figure}
\vspace{1em}
\begin{table}[ht]
\centering
\begin{tabular}{ccccccccc}
$m$ & scenario & $f_X$      & $\sigma(x)$  & $N_0$ & $N$ & $r_L$ & $r_R$ & $h$ \\
\hline
$m_1$ & Homoscedastic & $f^h_X$ & 6 & $6\cdot 10^6$ & 150 & 25 & 25 & 3\\
$m_2$ & Homoscedastic& $\calN(0,1)$ & 6 & $10^6$ & 150 & 25 & 25 & 3\\
$m_3$ & Heteroscedastic& $\calN(5,1)$ & $\frac{1}{6}e^{x/2}$&$10^6$ & 150 & 25 & 25 & 0.15
\end{tabular}
\caption{Settings and parameters for each mean function in the density estimation.}
\label{t:setting}
\end{table}

Figure \ref{f:ns:mean} depicts the mean functions $m$ and their corresponding PLR estimates.
The specific settings and parameters associated with each mean function are given in Table~\ref{t:setting}.
Based on these settings, the figure presents the results of obtaining $N=150$ data points sampled from the proposal PDF with both known $m$ through Algorithm~\ref{a:ISsampler} and PLR estimates via Algorithm~\ref{a:adaptive}. This offers insights into how the drawn samples are distributed. The red dashed lines in the figure indicate the thresholds $x_L$ and $x_R$.
For the heteroscedastic case, Figure~\ref{f:ns:mean} depicts the obtained samples of the variable $Y$ (labeled as ``m3 - Y") and the transformed variable $Z=\log(Y)$ (labeled as ``m3 - Z"), as discussed in Section~\ref{s:opt-htr}. The points for $Y$ are sampled as if the relation was homoscedastic.

Figure~\ref{f:ns:f1} compares various sampling strategies and the estimated density results under the settings given in Table~\ref{t:setting}, repeated 100 times. Black points (lines) represent the density estimates over the observed range from the smallest to largest value $Y_i$, while blue points correspond to the density estimates computed (extended) beyond this range.
The true log density values are marked by red lines, while green dashed lines indicate the $25$th smallest and largest $Y$ observations, which also serve as the thresholds for the GPD fitting when the modified estimator is used.
Throughout this section, we use the following terms in the labels to denote the distinct sampling strategies:
``random" represents results from random sampling of $Y$; ``optimal" and ``PLR" show results obtained using the optimal proposal PDF via Algorithms~\ref{a:ISsampler} and \ref{a:adaptive}, respectively; any label with ``modified" signifies the use of the GPD fit in the tail, as discussed in Section~\ref{s:gpd}.  In Figure~\ref{f:ns:f1}, labels with ``- N" or ``- N0" refer to the sample size used to compute the estimator. If not specifically indicated, the results are based on a sample size of $N$ observations.

For Figure~\ref{f:ns:f1m1}, which concerns the monotone function $m_1$, the following observations can be made:
\begin{itemize}
\item Using our proposal PDF in \eqref{e:mtd:g}, we considerably widen the observed sample range and the range where the target PDF is estimated reasonably well. This is also evident when contrasting the green dashed lines in ``random - N" and ``optimal - N" panels.

\item Even with a substantially larger sample size $N_0$, kernel density estimation is challenging in the tails due to data scarcity, as observed in ``random - N0". In regions with little or no data, the estimates tend to conform to the shape of the kernel, in our case Gaussian, which is parabolic on the log scale. For our kernel density estimation in \eqref{e:density-ks}, both the ``optimal" and ``PLR" estimates also take the Gaussian kernel shape in the far tails, particularly outside the observed range.

\item The modified estimator in \eqref{e:fyhatm} successfully recovers the distribution tail beyond the observed data, as in ``modified - N" or ``PLR modified - N" panels. From the true density curve for $m_1$, note a curvature change around $y=\pm 30$. For GPD fitting to work well, thresholds must be set beyond these points. A more detailed discussion on this can be found in Section~\ref{s:data:other:N0}.

\end{itemize}
For \( m_1 \), our optimal, modified, and PLR estimates approximate well the true density curve and accurately capture the curvature changes in the distribution tails. Further discussion on the optimality of the choice of \( p_X \) is postponed to Section~\ref{s:data:other:opt}.

Figure~\ref{f:ns:f1m2} provides results for the piecewise monotone function $m_2$. Many of the observations for Figure~\ref{f:ns:f1m1} apply for Figure~\ref{f:ns:f1m2} as well. Here, a noticeable curvature change occurs around $y=\pm 30$ for the true density curve. The GPD fits start beyond these thresholds, capturing the distribution tail.

Figure~\ref{f:ns:f1m3} presents results for the heteroscedastic scenario associated with $m_3$. We compare the ``random - N", ``optimal - N", and ``modified - N" estimators across two distinct schemes. The first-row panels, labeled ``m3", follow the approaches used in Figures~\ref{f:ns:f1m1} and~\ref{f:ns:f1m2}, treating the scenario as homoscedastic. In contrast, the second-row panels, denoted ``m3 - transformed", adhere to the procedures outlined in Section~\ref{s:opt-htr}. Here, we first transform the variable to \( Z=\log(Y) \), and subsequently estimate \( f_Z \) via \eqref{e:setting-2-density-ks}. The resulting estimators for \( f_Z \) are expected to exhibit similar behaviors seen in earlier homoscedastic cases. We then compute \( \widehat f_Y \) using \eqref{e:setting-2-densities-YZ}. The “optimal - N” density estimate showcases improved performance achieved through this transformation, especially evident in the right distribution tail. The ``modified - N" estimator also demonstrates its ability to capture the shape of the distribution tail.

\subsection{Discussion of other points} 
\label{s:data:other}
\subsubsection{Role of $N_0$ and usefulness of GPD}
\label{s:data:other:N0}
\begin{figure}[t]
\centering
\includegraphics[width=.98\linewidth]{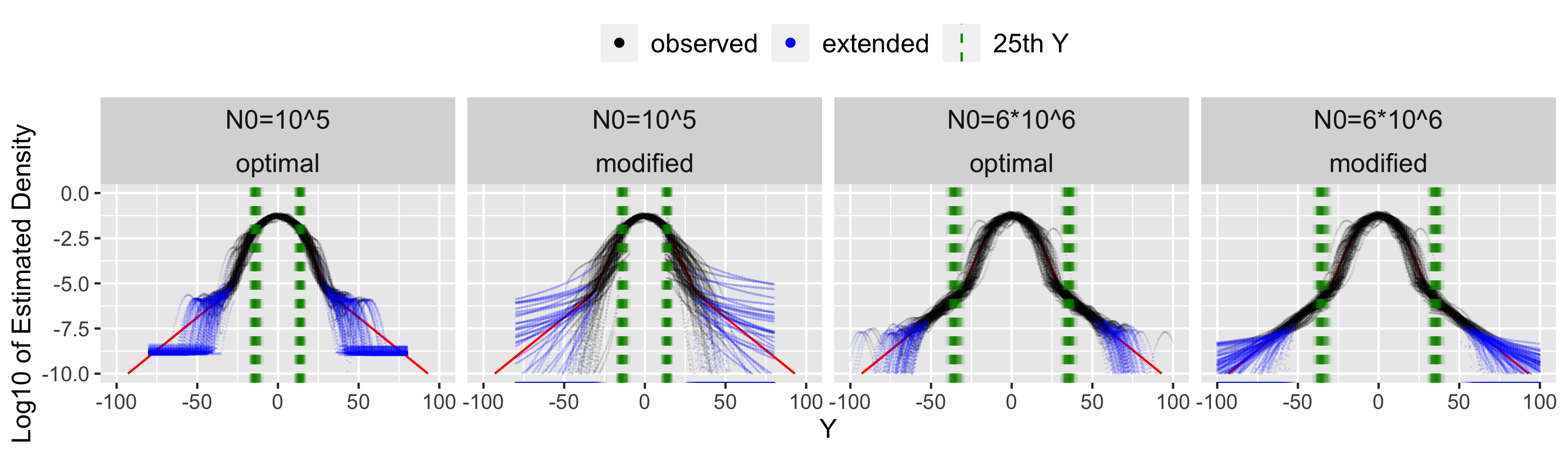}
\caption{ Comparison of estimated log-PDF for $m_1$ across different $N_0$ sizes. }
\label{f:ns:N}
\end{figure}

In our setup, \( N_0 \) needs to be chosen first. This parameter is important for two main reasons: firstly, it dictates the range where the target PDF could reliably be estimated; and secondly, it affects the GPD threshold and potential usefulness of GPD.
Figure~\ref{f:ns:N} compares the performance of ``optimal" and ``modified" results for \( m_1 \) with $N_0=10^5$ and $N_0=6\cdot 10^6$, while keeping $N=150$ in both scenarios.
The results show that a larger \( N_0 \) widens the range for reliable estimation.
Moreover, when examining the ``modified" results for the two $N_0$ values, it is evident that a smaller \( N_0 \) leads to GPD fitting for too small thresholds, failing to capture the curvature changes in the distribution tails. This indicates that GPD fitting with inadequate \( N_0 \) may not yield any benefits, as it does not accurately represent tail behavior.
Choosing a suitable threshold for GPD fitting is arguably a delicate issue that should ideally be based on the underlying ``physics" of the studied phenomenon (e.g. \cite{PIPIRAS2020}). 

\subsubsection{Optimality illustration}
\label{s:data:other:opt}
\begin{figure}
\centering
\begin{subfigure}{.98\linewidth}
    \centering
    \includegraphics[width=1\linewidth]{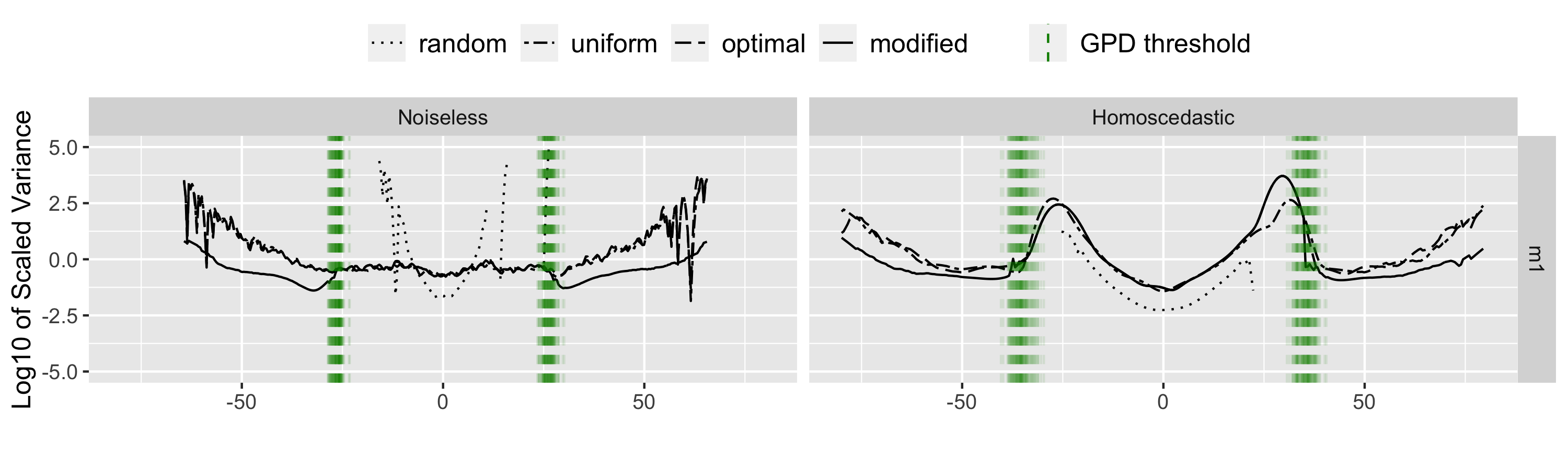}
    \vspace{-2.2em}
\end{subfigure}
\\ 
\begin{subfigure}{.98\linewidth}
    \centering
    \includegraphics[width=1\linewidth]{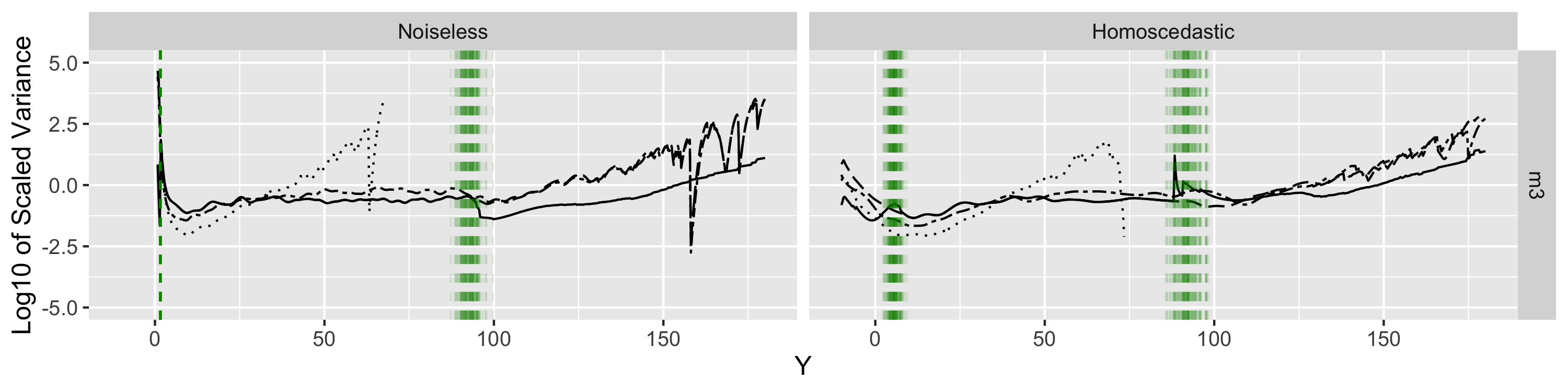}
     \vspace{-1em}
\end{subfigure}
\caption{ Log of scaled variance for $m_1$ (top) and $m_3$ (bottom) under noiseless (left) and homoscedastic (right) settings.}	
\label{f:ns:var}
\end{figure}

In Section~\ref{s:optimal}, we proposed the concept of optimality as described in \eqref{e:setting-0-optim}. Based on this definition, our optimal \( p_X \) was derived to ensure that the scaled variance of the density estimator is approximately constant within the GPD thresholds under the noiseless setting \eqref{e:setting-0}. 
Figure~\ref{f:ns:var} offers a visual illustration of this, showing the log of the empirical scaled variance for \( m_1 \) and \( m_3 \) under noiseless and homoscedastic scenarios. 
The settings for all scenarios are described in Table~\ref{t:setting2}. We note that $m_3$ is now used with homoscedastic errors, in contrast to Section~\ref{s:data:ex} which employed the heteroscedastic scenario.

\begin{figure}
\centering
\begin{subfigure}{.98\linewidth}
    \centering
    \includegraphics[width=1\linewidth]{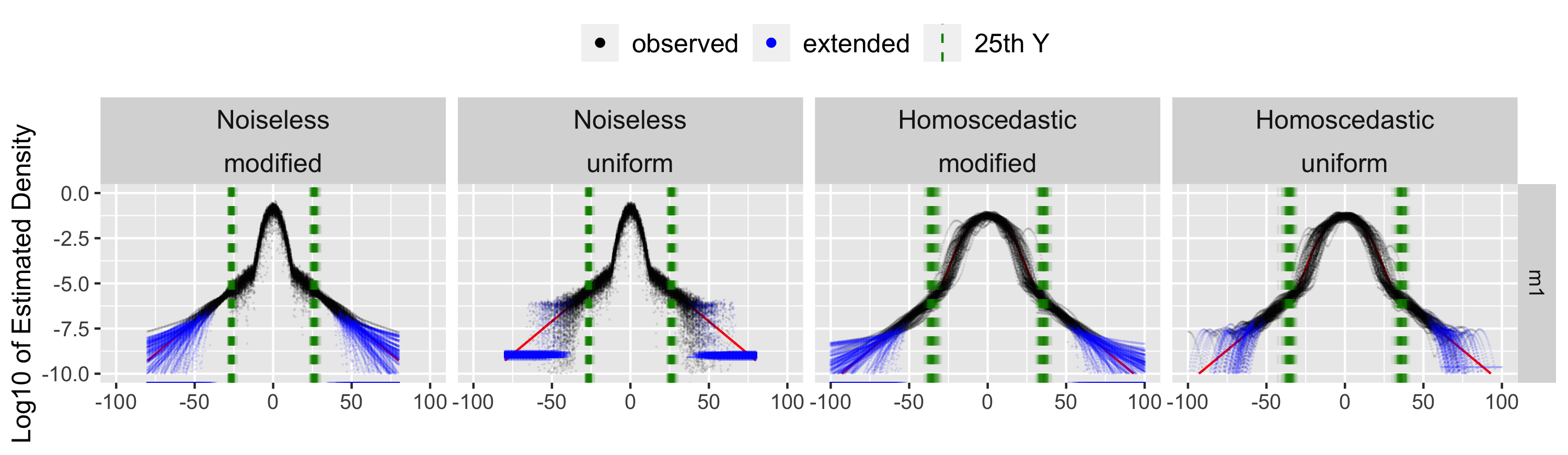}
    \vspace{-1.2em}
\end{subfigure}
\\ 
\begin{subfigure}{.98\linewidth}
    \centering
    \includegraphics[width=1\linewidth]{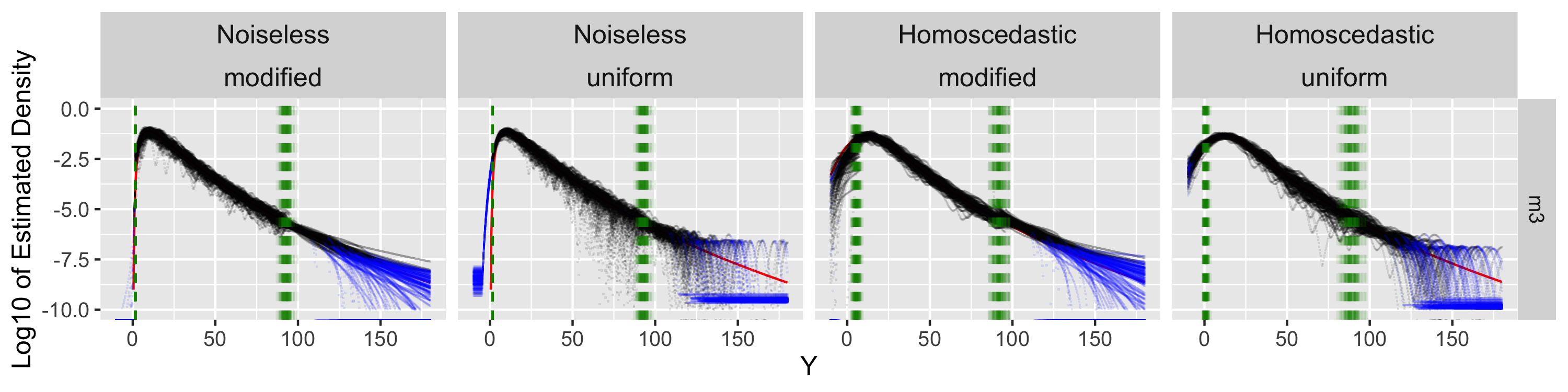}
     \vspace{-1em}
\end{subfigure}
\caption{ Comparison of estimated log-PDF for $m_1$ (top) and $m_3$ (bottom) between ``modified" and ``uniform" methods.}	
\label{f:ns:unif}
\end{figure}
\begin{table}[ht]
\centering
\begin{tabular}{ccccccccc}
$m$ & scenario &  $f_X$      & $\sigma(x)$  & $N_0$ & $N$ & $r_L$ & $r_R$ & $h$ \\
\hline
$m_1$ & Homoscedastic & $f_X^h$ & 6 & $6\cdot 10^6$ & 150 & 25 & 25 & 3\\
$m_1$ & Noiseless & $f_X^h$ & 0 & $10^6$ & 150 & 25 & 25 & 0.5\\
$m_3$ & Homoscedastic& $\calN(5,1)$ & 6 &$10^6$ & 150 & 25 & 25 & 3\\
$m_3$ & Noiseless & $\calN(5,1)$ & 0&$10^6$ & 150 & 25 & 25 & 1.5
\end{tabular}
\caption{Settings and parameters for each mean function in the optimality illustration.}
\label{t:setting2}
\end{table}
In Figure~\ref{f:ns:var}, the ``optimal" and ``modified" methods yield identical estimators, represented by the black solid line, in the middle range within the GPD thresholds, marked by the green dashed lines. They diverge beyond the GPD thresholds, with the modified estimator exhibiting lower variance. To compare against the optimal $p_X$, we have included the case when $p_X$ is the uniform density on the interval $\xlxr$, which is labeled as ``uniform''. All curves are plotted only over the ranges where data are observed, since the estimates tend to be unreliable beyond this range, as shown in Figure~\ref{f:ns:f1}.
For the case of random sampling, labeled ``random", we note that the scaled variance is small around the center; however, increases rapidly moving away from the center. Compared to the ``random" case, our optimal \( p_X \) performs more consistently over a wider range, particularly in the distribution tails. In the noiseless setting, as expected from our optimality criterion, our results confirm that the scaled variance for the optimal proposal PDF remains approximately constant within the GPD thresholds.

When comparing the ``uniform" and ``optimal" approaches under homoscedastic noise settings, we first note that uniform sampling is the best strategy for linear relationships according to our optimality criterion (see Example~\ref{ex:setting-0-linear}). Consequently, we observed similar performance levels between the two approaches for the linear model $m_1$.
To better highlight the differences, we examined the exponential function \( m_3 \) under homoscedastic noise setting. As seen from Figure~\ref{f:ns:mean}, the exponential relationship with $m_3$ shows more evident nonlinearity, for which we expect our ``optimal" sampling strategy to be beneficial. Indeed, we observe that the ``optimal" (or equivalently ``modified") sampling strategy exhibits a lower scaled variance towards the right distribution tails compared to the ``uniform" approach within the green dashed lines. The density estimation results for these settings are also illustrated in Figure~\ref{f:ns:unif}, showcasing comparisons across ``modified" and ``uniform" approaches in both noiseless and homoscedastic scenarios for \( m_1 \) and \( m_3 \). The variability also appears visually smaller for the ``modified" approach.

\subsection{Application to ship motions}
\label{s:data:ship}

\begin{figure}
\centerline{
\includegraphics[width=3in,height=2.5in]{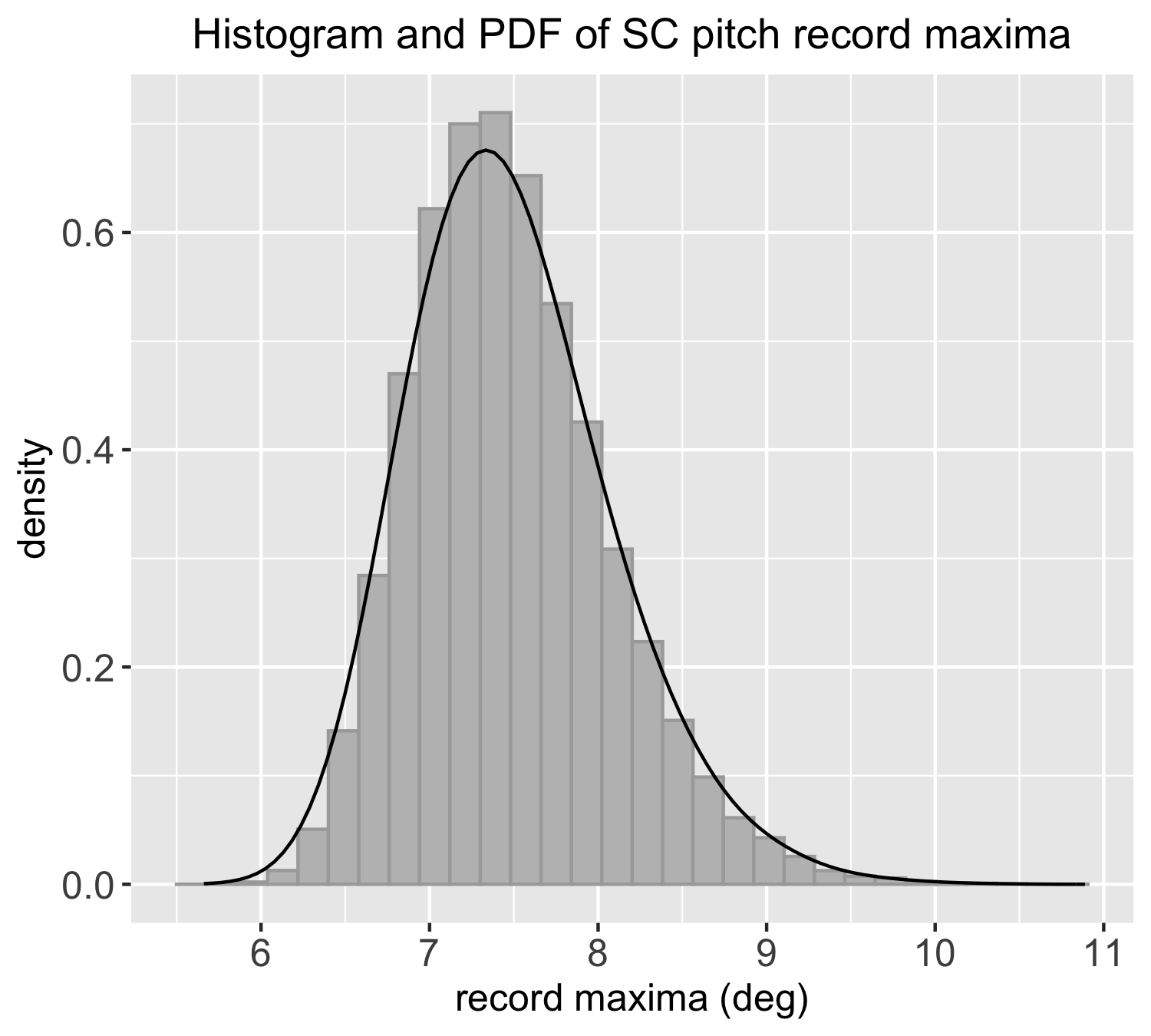}
\includegraphics[width=3in,height=2.5in]
{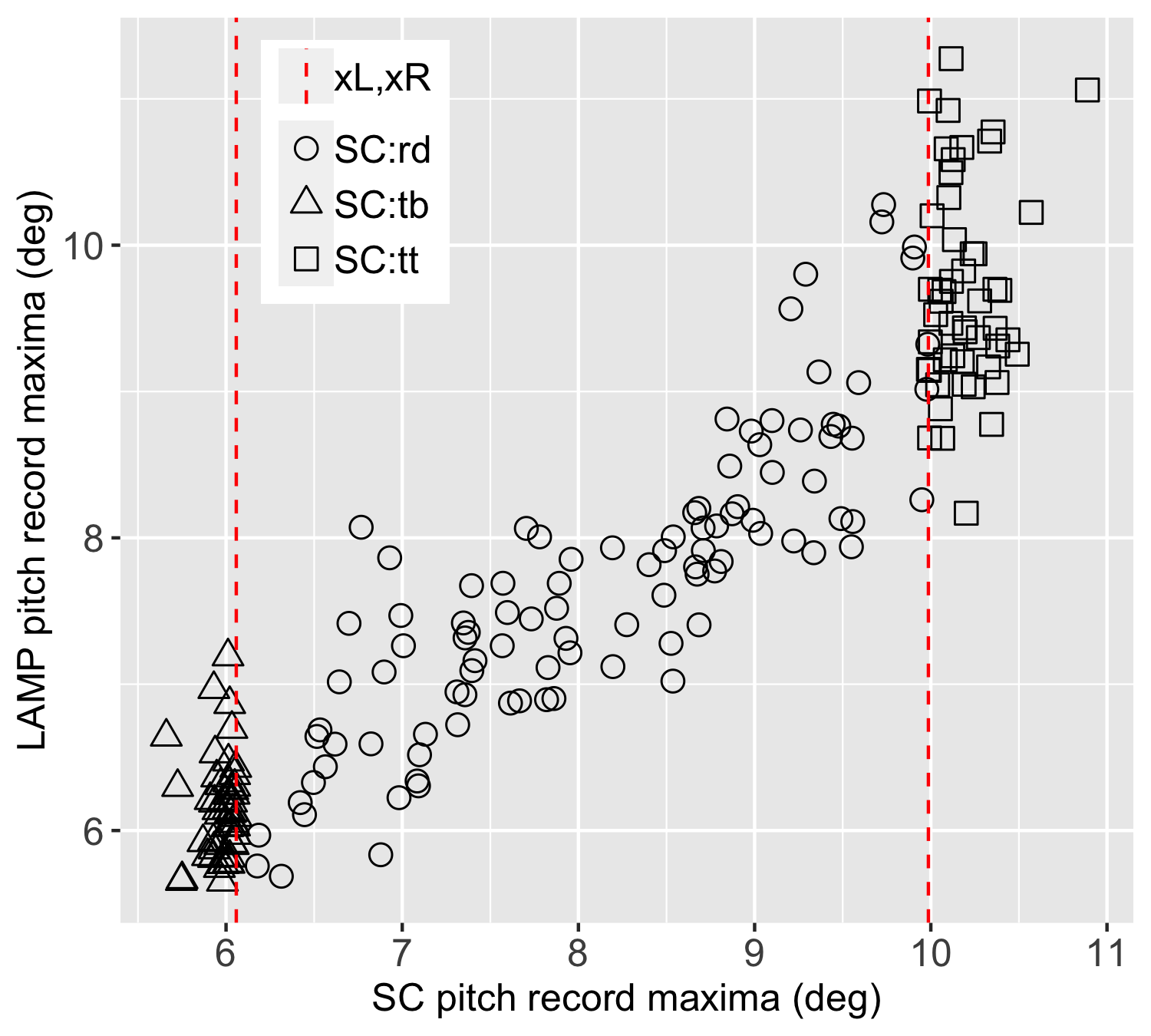}
}
\caption{Left: Histogram and PDF of SC pitch record maxima based on 100,000 observations. Right: The scatter plot of LAMP/SC pitch record maxima obtained via Algorithm~\ref{a:ISsampler} with ``uniform" $p_X$.}	
\label{f:ship:pitch}
\end{figure}

We illustrate here the considered approach in the ship motion application discussed in Section~\ref{s:introduction}. As in the left plot of Figure~\ref{f:intro-heave}, we focus on LAMP/SC ship motions but consider the pitch motion for the same ship in head seas, 10 kts speed and other conditions that are of little importance to understanding the illustration. We consider LAMP/SC pitch record maxima $Y/X$ and are interested in estimating the LAMP pitch record maximum PDF $f_Y(y)$. To apply the importace sampling approach, we first generate $N_0=100,000$ SC records. The histogram and estimated density of these SC pitch record maxima $X=\Xio,~ i=1,\dots, N_0$, are depicted in Figure~\ref{f:ship:pitch}, left plot. This PDF is estimated using kernel smoothing with a bandwidth of $0.2$.

Having the estimate of the PDF $\fx$, we need to decide on the proposal PDF $\px$.
In general, Algorithm~\ref{a:adaptive} can be employed for incorporating both mean function estimation and sampling. In our specific application, we proceed with a uniform proposal PDF between $r_L=50$ smallest and $r_R=50$ largest values $\xlxr$ (notably optimal when the mean function is linear). The proposal PDF is used to choose $N=200$ SC records and generate the associated LAMP record values. The right plot of Figure~\ref{f:ship:pitch} depicts a scatter plot of the sampled values, obtained using Algorithm~\ref{a:ISsampler}. This plot shows a (roughly) linear relationship between LAMP and SC outputs, supporting our choice of the uniform PDF on the interval $\xlxr$.

\begin{figure}[tb]
\centering    \includegraphics[width=.75\linewidth]{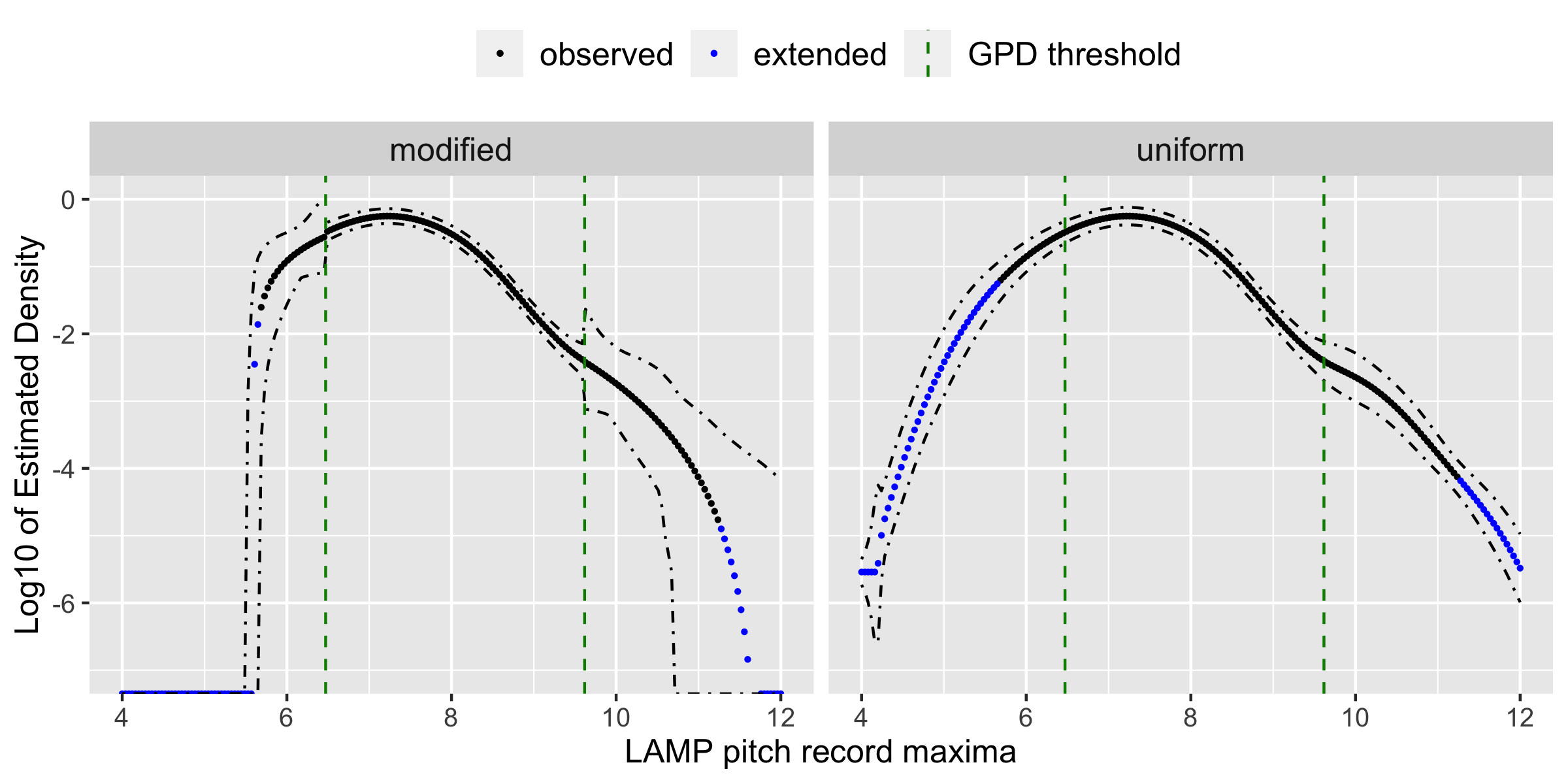}
\caption{Log of estimated PDF $\fy$ with approximated $95\%$ confidence interval when using ``modified" and ``uniform" approaches.}
\label{f:ns:ship:density}
\end{figure}
We are now equipped to estimate the target PDF $\fy$ through the importance sampling estimator \eqref{e:density-ks} and the modified estimator \eqref{e:fyhatm}. The resulting density estimates are presented in Figure~\ref{f:ns:ship:density}, labeled as ``uniform" and ``modified", respectively. As in the simulations above, the estimates beyond the data range are depicted in blue. For the kernel density estimates, we chose a bandwidth of $h=0.4$. Regarding the modified estimator, we set the GPD thresholds by selecting extreme observations among the $Y$ values: the left threshold is set at the 55th smallest, and the right threshold at the 30th largest observation. The estimated GPD parameters are $(\hxi_L, \hdel_L) = (-0.4766, 0.4192)$ for the left tail and $(\hxi_R, \hdel_R) = (-0.2104, 0.4464)$ for the right tail.

In addition to the density estimates, we have included approximate $95\%$ confidence intervals in the plot. To have non-negative density estimates, we employ the delta method for constructing confidence intervals on the log-transformed estimates. To be specific, for the kernel density estimate, we approximate the $100(1-\alpha)\%$ confidence interval for $\log \fyhat$ as
\begin{equation}\label{e:ship:CI:KS}
    \log \fyhat \pm z_{1-\frac{\alpha}{2}}\sqrt{\frac{\widehat{\var}(\fyhat)}{\fyhat^2}},
\end{equation}
where $z_{1-\frac{\alpha}{2}} = \Phi^{-1}(1-\frac{\alpha}{2})$ corresponds the upper $1-\frac{\alpha}{2}$ percentile of the standard normal distribution, and $\widehat{\var}(\fyhat)$ is obtained based on \eqref{e:density-ks-var}. The confidence interval for $\fyhat$ is then obtained by exponentiating this interval.

When implementing the modified estimator, the central part of the distribution employs the kernel density estimate, and we use \eqref{e:ship:CI:KS} for the confidence interval.
However, for the tails of the distribution, the density of a given target above the threshold is defined by the product of the probability of exceeding a threshold \eqref{e:gpd:cL} and the PDF of GPD \eqref{e:gpd}. For instance, the right tail estimate is $\widehat c_R' \cdot g_{\hxi_R, \hdel_R}(y-y_R)$ in \eqref{e:fyhatm}.
Assuming these two values are independent, the $100(1-\alpha)\%$ confidence interval for this term is derived by calculating the $100\sqrt{1-\alpha}\%$ confidence interval for each component term and multiplying the respective endpoints of these intervals.

Similar to the variance calculation for $\fyhat$ in \eqref{e:density-ks-var}, the variance of $\widehat c_R'$ is given by
\begin{equation}\label{e:ship-cR-var}
	\mbox{\rm Var}(\widehat c_R') = \frac{1}{N} \mbox{\rm Var}( \bbone(Y>y_R) w(X))
	= \frac{1}{N} \E ( \bbone(Y>y_R) w(X)^2) - \frac{1}{N} (\E \bbone(Y>y_R) w(X))^2.
\end{equation}
The $100\sqrt{1-\alpha}\%$ confidence interval for $\widehat c_R'$ is approximated as
\begin{equation}\label{e:ship:CI:gpd}
    \exp\left(\log\widehat c_R' \pm z_{\frac{1+\sqrt{1-\alpha}}{2}}\sqrt{\frac{\widehat{\var}(\widehat c_R')}{\widehat c_R'^2}}\right),
\end{equation}
where $\widehat{\var}(\widehat c_R')$ is estimated based on \eqref{e:ship-cR-var} using empirical quantities.
On the other hand, for the PDF of GPD, ML estimators $\hxi$ and $\hdel$ are computed from the sample $Y_1, \dots, Y_r$, which consists of $r$ observations exceeding the specified threshold. According to \cite{smithGPD}, the large sample asymptotics of the ML estimators are given by
\begin{equation}\label{e:ship:asymp}
    \sqrt{r}\left( \begin{array}{c}
        \hxi - \xi_0\\
        \hdel - \beta_0 
    \end{array} \right) \overset{d}{\rightarrow} \calN(0, W^{-1}),
\end{equation}
where $\xi_0$ and $\beta_0$ are the true values and 
\begin{equation}\label{e:ship:asymp:var}
    W^{-1} = \left( \begin{array}{cc}
        1+\xi_0 & -\beta_0 \\
        -\beta_0 & 2\beta_0^2
    \end{array} \right).
\end{equation}
We note that \eqref{e:ship:asymp} holds only when $\xi > -1/2$. In our practical application, we adjusted the choice of GPD threholds to ensure the parameters satisfy this condition. Once the ML estimators $(\hxi, \hdel)$ are obtained, we independently draw 100 data points, $(\hxi_b, \hdel_b), ~b=1,\dots, 100$, from the asymptotic distribution \eqref{e:ship:asymp}, replacing the true values $\xi_0$ and $\beta_0$ with $\hxi$ and $\hdel$. Then, in this parametric bootstrap approach, the $100\sqrt{1-\alpha}\%$ confidence interval is approximated through the sample quantiles of $g_{\hxi_b,\hdel_b}(y), ~b=1,\dots, 100$.
For other methods to set confidence intervals, see \cite{glotzer2017confidence}.

While the estimates for the ``modified" and ``uniform" approaches coincide between the GPD thresholds, we note from Figure~\ref{f:ship:pitch} that they are quite different in the distribution tails. The ``modified" estimate, in particular, suggests lighter tails than the ``uniform" estimate. As discussed in Section~\ref{s:est-t-k}, we would not rely on the latter approach beyond observed data depicted in blue in the figure.


\section{Conclusions}
\label{s:conclusions}

In this work, we proposed an importance sampling framework for choosing low-fidelity outputs to generate the corresponding high-fidelity outputs and to estimate their PDF, with the emphasis on the tails. At the center of our analysis lied the notion of optimal proposal PDF for importance sampling. The proposed approach performed well in simulations and was illustrated on an application. 

Several problems related to this work could be studied in the future. 
We noted in Section~\ref{s:introduction} that other approaches would seek importance sampling schemes for the underlying random components of the system of interest, that is, the variable $\varepsilon_n, ~ n=1,\dots,N_w$, in \eqref{e:intro:Longuet-height} for our application. In higher dimensions (large $N_w$ as in our application), this is a challenging problem and when approaches to tackle it become better developed, our method should be compared to them in terms of performance.
Moreover, incorporating costs of low- and high-fidelity outputs and considering more than two sources of data, as explored by \cite{pham2022ensemble}, \cite{jakeman2022adaptive}, \cite{han2023approximate}, and others, present another interesting direction.


\section*{Acknowledgments} 
This work was supported in part by the ONR grants N00014-19-1-2092 and N00014-23-1–2176 under Dr.\ Woei-Min Lin. The authors are grateful to Dr.\ Arthur Reed at NSWC Carderock Division for discussions that led to the formulation of and the approaches to the problem addressed in this work. The authors also thank Dr.\ Vadim Belenky at NSWC Carderock Division for comments on this work.
Finally, the authors are grateful to two anonymous Reviewers whose comments helped improving the paper considerably.

\small
\bibliography{mf-sampling-rev}

\flushleft
\begin{tabular}{lll}
Minji Kim, Vladas Pipiras &Kevin O'Connor & Themistoklis Sapsis\\
Dept.\ of Statistics and Operations Research & & Dept.\ of Mechanical Engineering\\
UNC at Chapel Hill &  Optiver & Massachusetts Institute of Technology\\
CB\#3260, Hanes Hall & Suite \#800, 130 East Randolph & Room 5-318, 77 Massachusetts Av.\\
Chapel Hill, NC 27599, USA & Chicago, IL 60601, USA & Cambridge, MA 02139, USA\\
{\it mkim5@unc.edu, pipiras@email.unc.edu}  & {\it oconnor.kevin.ant@gmail.com} & {\it sapsis@mit.edu} \\
\end{tabular}

\end{document}